%% file: paper.tex
\documentclass[12pt]{article}

\usepackage{amsmath}
\usepackage{amsfonts}
\usepackage{amssymb}

\usepackage{graphicx}
\usepackage{epsfig}
\usepackage{epsf}

\allowdisplaybreaks

\begin{document}

\author{V. Cuplov ${}^\dagger$ \and A. Nehme ${}^\ddagger$}
  
\title{Isospin Breaking in $K_{\ell 4}$ Decays of the \\
       Charged Kaon}

\maketitle

\begin{center}
${}^\dagger$ Universit\'e de la M\'editerran\'ee \hfill ${}^\ddagger$ Universit\'e Sainte Marthe \\
Centre de Physique Th\'eorique \hfill D\'epartement de Physique \\ 
171 avenue de Luminy \hfill 74 rue Louis Pasteur \\ 
F-13288 Marseille Cedex 9 \hfill F-84029 Avignon Cedex 1 \\
\texttt{cuplov@cpt.univ-mrs.fr} \hfill \texttt{Abass.Nehme@univ-avignon.fr}
\end{center}

\begin{abstract}
The charged $K_{\ell 4}$ decay, $K^+\rightarrow\pi^+\pi^-\ell^+\nu_{\ell}$ is studied in the framework of chiral perturbation theory based on the effective Lagrangian including mesons, photons, and leptons. We give analytic expressions for the two vectorial form factors, $f$ and $g$, calculated at one-loop level in the presence of Isospin breaking effects. These expressions may then be used to disentangle the Isospin breaking part from the measured form factors and hence improve the accuracy in the determination of $\pi\pi$ scattering parameters from $K_{\ell 4}$ experiments.
\end{abstract}

\textbf{keywords:} Electromagnetic Corrections, Kaon Semileptonic Decay, Form Factors, Chiral Perturbation Theory.

\pagebreak 

\tableofcontents

\input{introduction}

\input{section_1}

\input{section_2}

\input{section_3}

\input{conclusion}

\begin{flushleft}
\textit{\textbf{Acknowledgements}}

We express our gratitude to Marc Knecht for suggesting the study, for his encouragement and help. 
\end{flushleft} 

\bibliographystyle{unsrt}

\bibliography{paper}

\begin{figure}[p]
\epsfxsize14cm \centerline{\epsffile{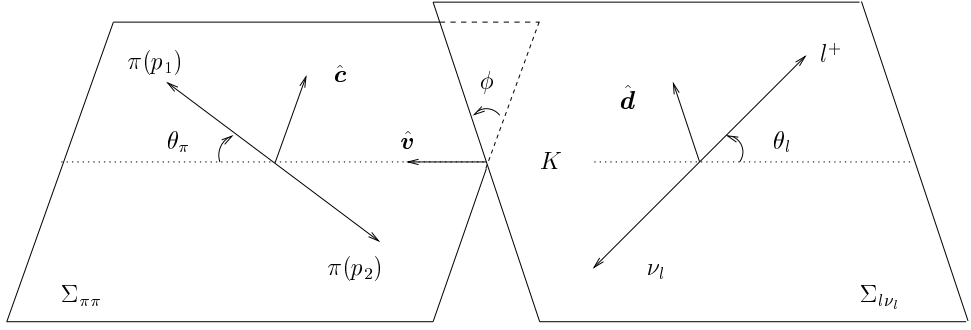}}
\caption{\label{fig:kinematics} Angles and unit vectors used in the
kinematical description of $K_{\ell 4}$ decays. $\Sigma_{\pi\pi}$ and
$\Sigma_{l\nu_l}$ are the planes defined in the kaon rest frame by
the pion pair and the lepton pair, respectively. $\theta_{\pi}$
($\theta_l$), the angle formed by $\boldsymbol{p}_1$
($\boldsymbol{p}_l$), in the dipion (dilepton) rest frame, and the
line of flight of the dipion (dilepton) as defined in the kaon
rest frame. $\phi$, the angle between the normals to
$\Sigma_{\pi\pi}$ and $\Sigma_{l\nu_l}$. $\hat{\boldsymbol{v}}$ is
a unit vector along the direction of flight of the dipion in the
kaon rest frame. $\hat{\boldsymbol{c}}$ ($\hat{\boldsymbol{d}}$)
is a unit vector along the projection of $\boldsymbol{p}_1$
($\boldsymbol{p}_l$) perpendicular to $\hat{\boldsymbol{v}}$.}
\end{figure}

\begin{figure*}[p]
\epsfxsize14cm \centerline{\epsffile{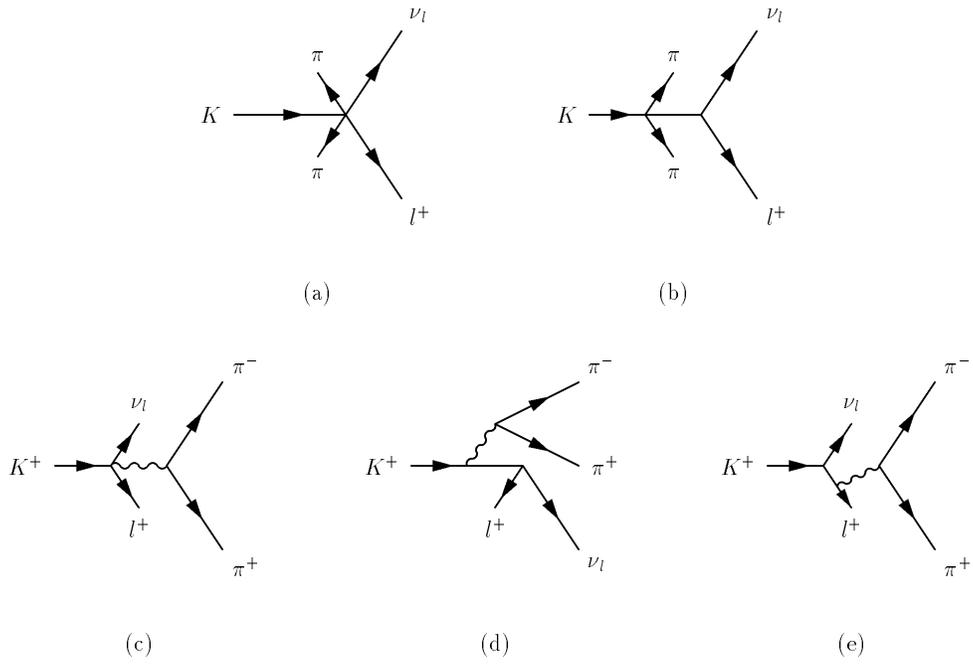}}
\caption{\label{fig:tree} Feynman diagrams representing $K_{\ell 4}$ decay
amplitudes at tree level. Wavy lines stand for photons. Only
diagrams (a) and (b) contribute to the decay amplitudes ${\cal
A}^{00}$ and ${\cal A}^{0-}$ corresponding to $K_{\ell 4}$ decays of the charged kaon to neutral pions and of the neutral kaon to a neutral and a charged pion, respectively.}
\end{figure*}

\begin{figure}[p]
\epsfxsize14cm \centerline{\epsffile{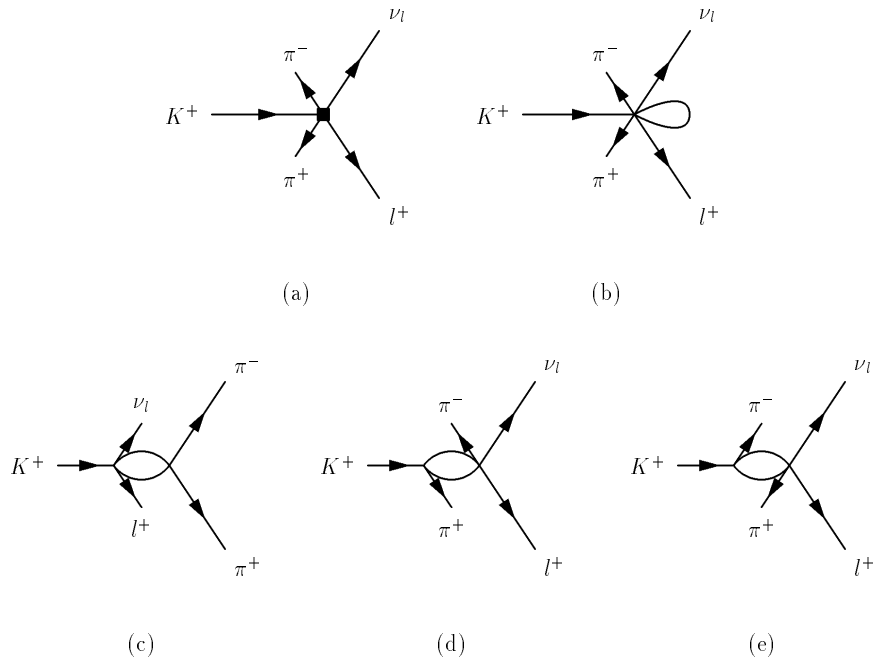}}
\caption{\label{fig:strong} Feynman diagrams representing the $K_{\ell 4}$ decay amplitude of the charged kaon at one-loop. Is shown only the non photonic topology. Diagram (a) represents Born and counter-terms contributions. Tadpole contribution is given by diagram (b). Diagrams (c), (d) and (e) stand for contributions from the $s$-, $t$- and $u$-channels, respectively.}
\end{figure}

\begin{figure}[p]
\epsfxsize14cm \centerline{\epsffile{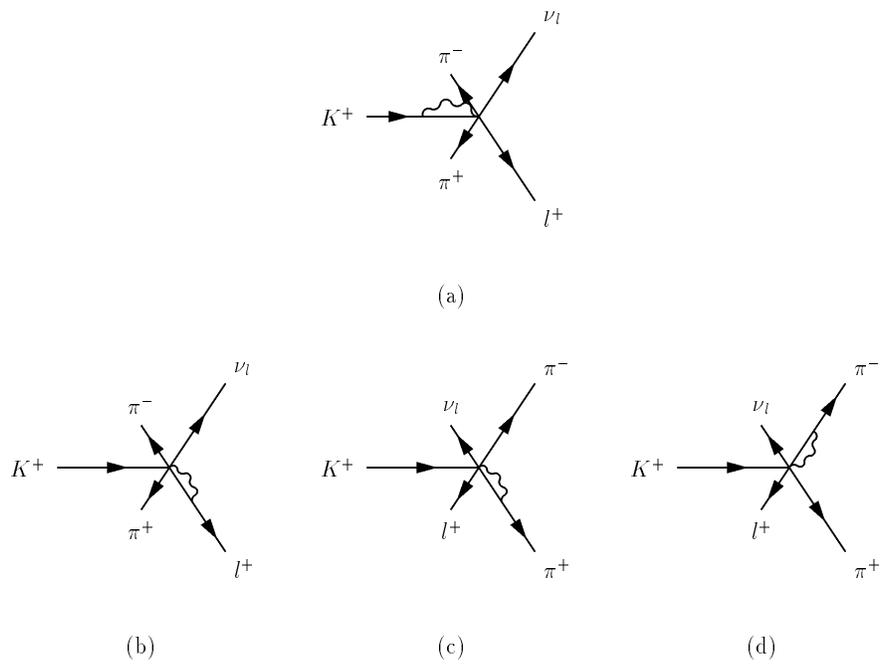}}
\caption{\label{fig:vertex-leg} Virtual photons in Feynman diagrams for the $K_{\ell 4}$ decay in the charged channel. The topology consists on attaching a photon on the pure vertex from one side and a leg from the other side.}
\end{figure}

\begin{figure}[p]
\epsfxsize14cm \centerline{\epsffile{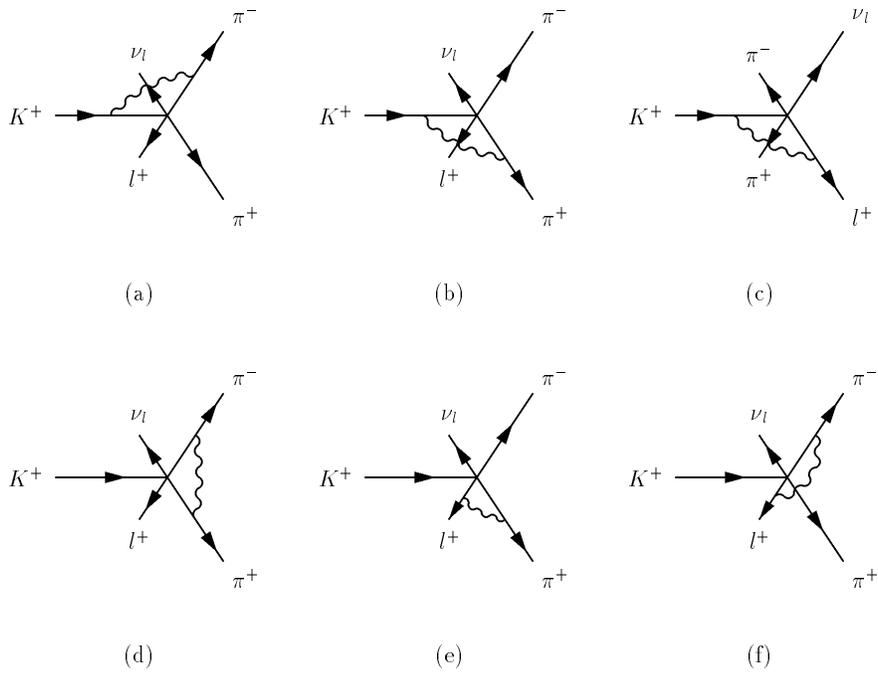}}
\caption{\label{fig:leg-leg} Virtual photons in Feynman diagrams for the $K_{\ell 4}$ decay in the charged channel. The topology consists on attaching a photon between two legs of the pure vertex.}
\end{figure}

\begin{figure}[p]
\epsfxsize14cm \centerline{\epsffile{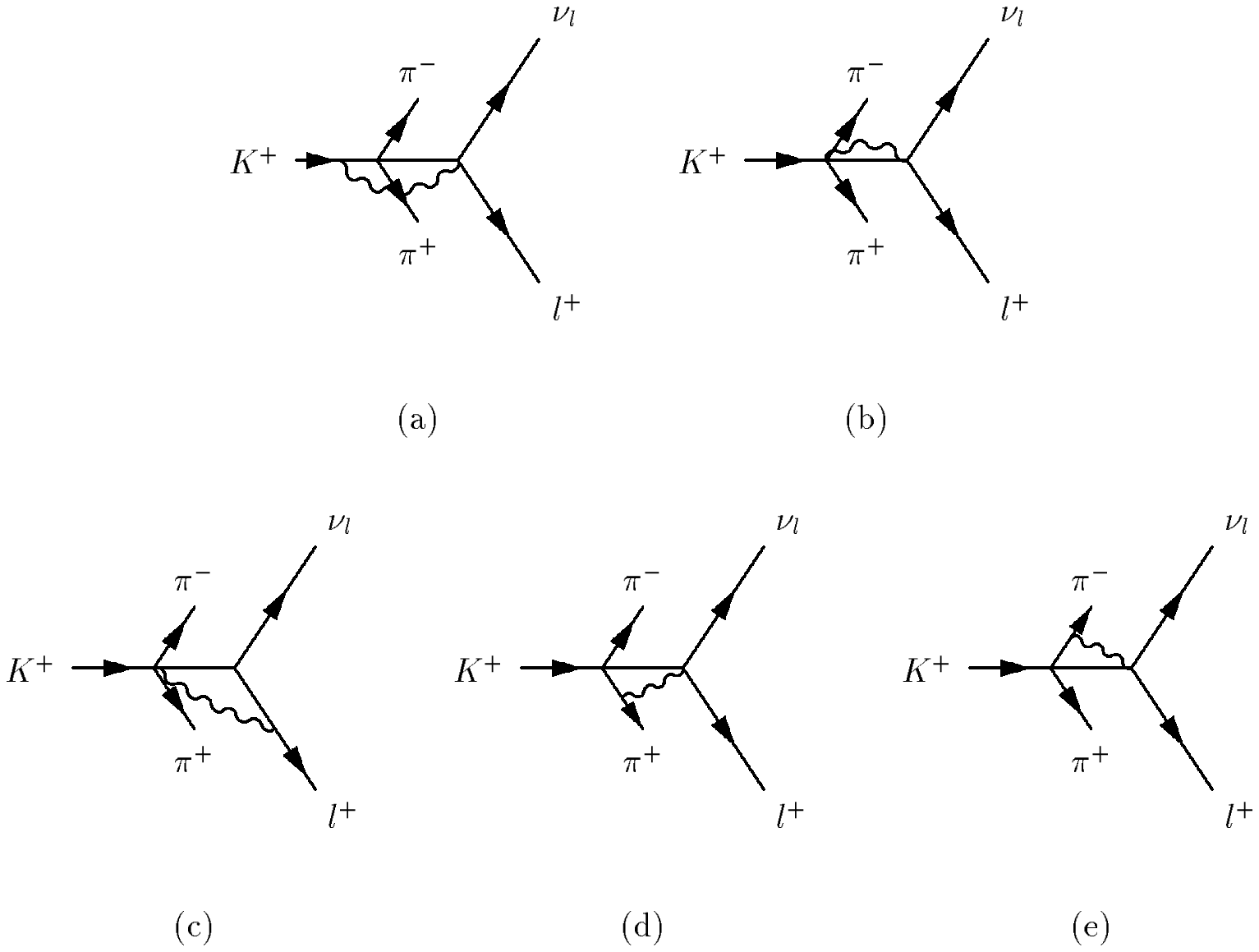}}
\caption{\label{fig:R-vertex-leg} Virtual photons in Feynman diagrams for the $K_{\ell 4}$ decay in the charged channel. One takes the pure vertex mediated by a $K^+$ pole and then attaches a photon between a vertex and a leg.}
\end{figure}

\begin{figure}[p]
\epsfxsize14cm \centerline{\epsffile{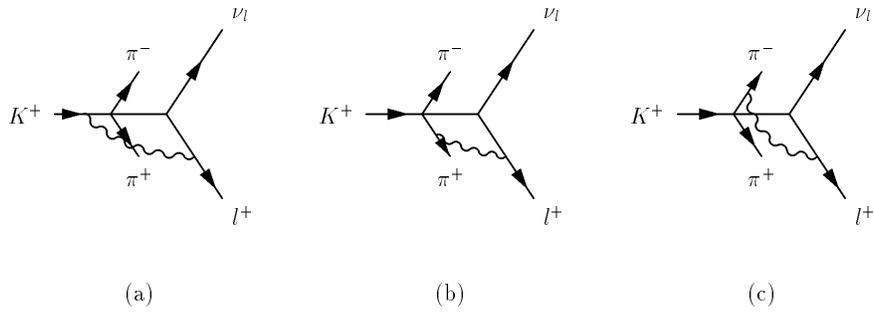}}
\caption{\label{fig:R-leg-leg} Virtual photons in Feynman diagrams for the $K_{\ell 4}$ decay in the charged channel. One takes the pure vertex mediated by a $K^+$ pole and then attaches a photon between two legs.}
\end{figure}

\begin{figure}[p]
\epsfxsize14cm \centerline{\epsffile{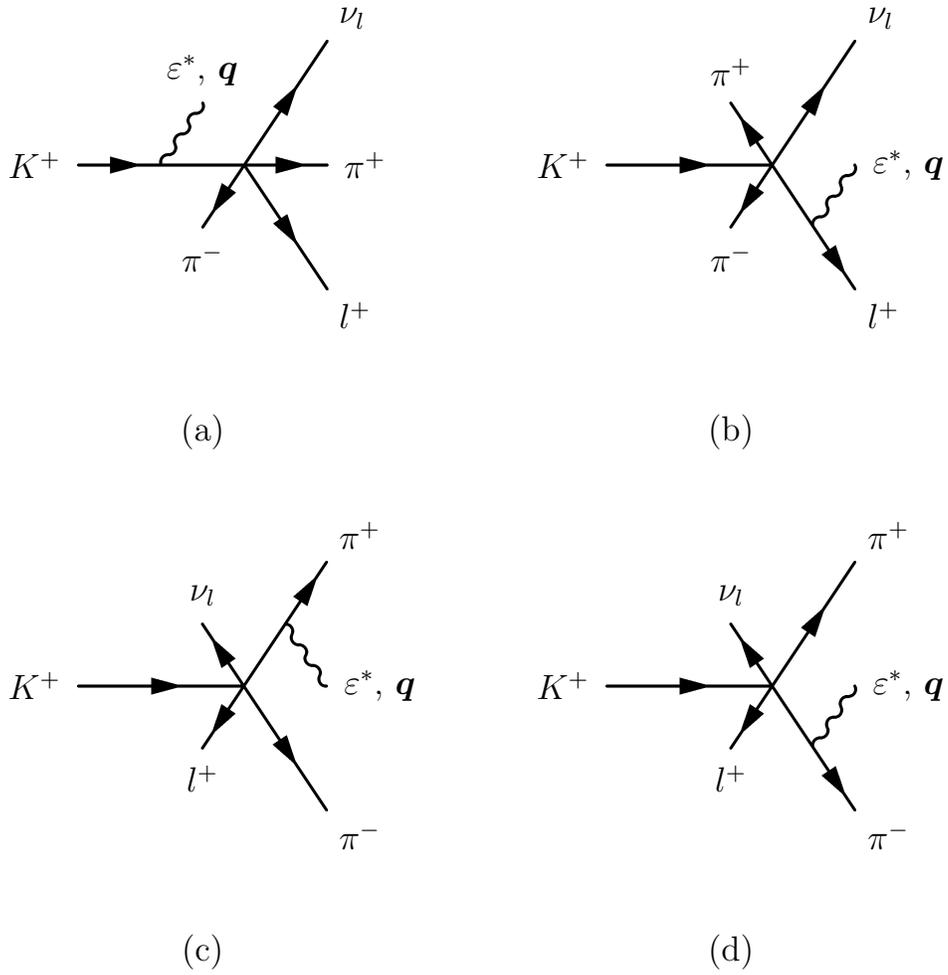}}
\caption{\label{fig:soft} Feynman diagrams representing the
contribution of $F$ and $G$ form factors to the bremsstrahlung
amplitude.}
\end{figure}

\begin{table}[p]
\begin{center}
\begin{tabular}{cc}
\hline
\textbf{diagram} & $\delta F\,=\,\delta G$ \\
\hline & \\
\textbf{3. (a)} & $-\displaystyle\frac{e^2}{4\pi^2}\,\ln m_{\gamma}^2$ \\ & \\
\textbf{5. (a)} & $-\displaystyle\frac{e^2}{8\pi^2}\,p\cdot p_2\tau
(p,p_2,M_K,M_{\pi})\ln
m_{\gamma}^2$ \\ & \\
\textbf{5. (b)} & $\displaystyle\frac{e^2}{8\pi^2}\,p\cdot p_1\tau
(p,p_1,M_K,M_{\pi})\ln m_{\gamma}^2$ \\ & \\
\textbf{5. (c)} &
$\displaystyle\frac{e^2}{8\pi^2}\,p\cdot p_l\tau (p,p_l,M_K,m_l)\ln m_{\gamma}^2$ \\ & \\
\textbf{5. (d)} & $-\displaystyle\frac{e^2}{8\pi^2}\,p_1\cdot p_2\tau
(p_1,-p_2,M_{\pi},M_{\pi})\ln m_{\gamma}^2$ \\ & \\
\textbf{5. (e)} & $\displaystyle\frac{e^2}{8\pi^2}\,p_1\cdot p_l\tau
(p_1,-p_l,M_{\pi},m_l)\ln m_{\gamma}^2$ \\ & \\
\textbf{5. (f)} & $-\displaystyle\frac{e^2}{8\pi^2}\,p_2\cdot p_l\tau
(p_2,-p_l,M_{\pi},m_l)\ln
        m_{\gamma}^2$ \\ & \\
\hline
\end{tabular}
\end{center}
\caption{\label{tab:infrared} Infrared divergent part of the corrected
$f$ and $g$ form factors due to virtual photon corrections. The
contribution from \textbf{diagram 3. (a)} comes from wave function
renormalization of external charged particles, $K^+$, $\pi^+$, $\pi^-$, and $\ell^+$.}
\end{table}

\end{document}

%% file: introduction.tex
\section{Introduction}
\label{introduction}

Nowadays, it is widely recognized that $\pi\pi$ scattering is the
purest process allowing a direct probe of the Quantum
Chromodynamics (QCD) ground state. This is due in part to the
\textit{fact} that pions are the lightest pseudo Goldstone bosons
coupled to the vacuum thanks to the mechanism of spontaneous and
explicit breaking of Chiral symmetry. Let $N_f$ be the number of
light quark flavors and denote by $F_0$ the coupling in question,
then, the Goldstone's, $\phi^a$, satisfy,
\begin{equation}
\langle 0|A_{\mu}^a(0)|\phi^b(p)\rangle\,=\,i\delta^{ab}F_0p_{\mu}\,, \quad
a,b=1,\ldots ,N_f^2-1\,,
\end{equation}
where $A_{\mu}^a$ are the usual axial currents.

The coupling $F_0$ is given by the pion decay constant in the limit of vanishing masses for the $N_f$ light quarks,
\begin{equation}
F_{\pi}\,=\,F_0\left[ 1+{\cal O}(m_q)\right ]\,, \qquad q=u,d,s\,.
\end{equation}
The non vanishing of $F_0$ is a necessary and sufficient condition for the spontaneous breaking of Chiral symmetry~\cite{Stern:1998dy}. While the occurrence of the latter mechanism is a fact, the way it occurs remains uncertain. It is currently \textit{assumed} that, resting on large-$N_c$ power counting grounds, the present mechanism is due to important quark condensation in the vacuum. To confirm or reject this assumption one has to \textit{measure} the quark condensate, $\langle{\overline q}q\rangle$. To this end, it is convenient to consider the ratio~\cite{Gell-Mann:1968rz},
\begin{equation}
\label{eq:flavor_ratio}
X\,\doteq\,-\frac{(m_u+m_d)\langle{\overline q}q\rangle}{F_{\pi}^2M_{\pi}^2}\,,
\end{equation}
which vanishes for a vanishing quark condensate and equals one for large value of $\langle{\overline q}q\rangle$. In order to fix the value for $X$ from experiment, it is interesting to perform accurate measurements of observables that are sensitive to
variations of $X$. This is the case of low-energy $\pi\pi$
scattering. In fact, the latter is solely described in the
threshold region in terms of the $S$-wave scattering lengths,
\begin{eqnarray}
a_0^0
&=& \frac{M_{\pi}^2}{96\pi F_{\pi}^2}\left[ 5\alpha +16\beta +\mathcal{O}(m_q)\right] \,,
\\
a_0^2
&=& \frac{M_{\pi}^2}{48\pi F_{\pi}^2}\left[ \alpha -4\beta +\mathcal{O}(m_q)\right] \,,
\end{eqnarray}
where,
\begin{equation}
\alpha\,=\,4-3X+\mathcal{O}(m_q)\,, \quad \beta\,=\,1+\mathcal{O}(m_q)\,.
\end{equation}
For instance, the isoscalar scattering length,
$a_0^0$, varies by a factor $2$ when $X$ varies from $1$ to zero.
This sensitivity of $\pi\pi$ scattering to the size of the quark
condensate is not only a feature of the leading order but, on the
contrary, it persists at higher orders allowing $\pi\pi$
scattering to be the golden process for testing the mechanism of
quark condensation~\cite{Knecht:1995tr}.

The $\pi\pi$ scattering is experimentally
accessible in $K_{\ell 4}$ and pionium decays. In fact, the partial wave
expansion of $K_{\ell 4}$ form factors displays the $\pi\pi$ phase
shifts as stated by the Watson final state interaction theorem~\cite{Watson:1952ji, Watson:1954uc}. The data on phase shifts can then be translated into a
model-independent determination of scattering lengths by means of
Roy equations~\cite{Ananthanarayan:2000ht}. Concerning the pionium characteristics such as its
lifetime, $\tau$, and $2S-2P$ strong energy level shift, $\Delta E_s$, they give direct access to $\pi\pi$ scattering lengths via~\cite{Efimov:1986},
\begin{equation}
\label{eq:pionium}
\tau\,\propto\,(a_0^0-a_0^2)^2\,, \quad \Delta
E_s\,\propto\,2a_0^0+a_0^2\,.
\end{equation}
Once the results from the
presently running DIRAC experiment are available, $a_0^0$ and
$a_0^2$ should be determined with $5$ to $10\%$ accuracy~\cite{Adeva:2003up}. On the
other hand, the charged $K_{\mathrm{e}4}$ decay has been measured
by the E$865$ experiment~\cite{Pislak:2001bf, Pislak:2003sv} and the outgoing data have been analyzed in~\cite{Colangelo:2001sp} and~\cite{Descotes-Genon:2001tn},
independently. Before recalling the conclusions of both
references, let us stress that the obtained value for $a_0^0$ is
$7\%$ accuracy and is compatible with the prediction of the
standard power counting which rests on large $N_c$ grounds. The
analysis of Ref.~\cite{Colangelo:2001sp} relies on chiral symmetry
constraints which correlate the two scattering lengths. It yields
a value for $a_0^2$ consistent with the standard counting
prediction. Furthermore, if one combines the extracted values for
the scattering lengths, the ratio (\ref{eq:flavor_ratio}) reads then,
$X\sim 0.94$. As for the analysis performed in
Ref.~\cite{Descotes-Genon:2001tn}, it combines the data of the E$865$
experiment with an existing one in the isospin-two channel below
$800$ MeV without using the aforementioned chiral constraints. The
conclusions of this analysis point out a discrepancy at the
$1-\sigma$ level between the extracted value for $a_0^2$ and the
one predicted by the standard power counting. Moreover, the
corresponding value for the ratio (\ref{eq:flavor_ratio}) was found to be,
$X\sim 0.81$. In view of the ``disagreement'' between the results of
the two analysis, and before drawing off any conclusion about the
size of the quark condensate, it is necessary to devote much more
effort at both the experimental and the theoretical levels. In
this direction, new precise measurements of charged and neutral $K_{\mathrm{e}4}$ decay are currently taking data at CERN~\cite{Batley:2000} and FNAL~\cite{Santos:2003}, respectively. This should be accompanied by an improvement of the accuracy in the theoretical prediction for the
scattering lengths by evaluating isospin breaking effects in
$\pi\pi$ scattering as well as in $K_{\ell 4}$ decays. While such
effects in the former case are now under control at leading and
next-to-leading orders~\cite{Maltman:1997nw, Maltman:1997nwE, Meissner:1997fa, Meissner:1997faE, Knecht:1998jw, Knecht:2002gz}, we are interested in evaluating these effects at the same orders for the latter case. Recently, one of us has published the calculation of Isospin breaking effects in the $K_{\ell 4}$ decay of the neutral kaon~\cite{Nehme:2003bz}. The aim of the present work is to evaluate the same effects in the $K_{\ell 4}$ decay of the charged kaon.

%% file: section_1.tex
\section{$K_{\ell 4}$ decay of the charged kaon}

The semileptonic $K_{\ell 4}$ decay of the charged kaon is given schematically by, 
\begin{equation}
\label{eq:process} 
K^+(p)\longrightarrow\pi^+(p_1)\,\pi^-(p_2)\,\ell^+(p_l)\,\nu_{\ell}(p_{\nu})\,, 
\end{equation} 
where the lepton, $\ell$, is either a muon, $\mu$, or an electron, e, and
$\nu$ stands for the corresponding neutrino.

\subsection{Matrix element}

The decay (\ref{eq:process}) is described in terms of an invariant
decay amplitude, ${\cal A}^{+-}$, defined via the matrix element,
\begin{eqnarray}
& & \langle\pi^+(p_1)\pi^-(p_2)\ell^+(p_l)\nu_{\ell}(p_{\nu})|K^+(p)\rangle
\nonumber \\ 
& & \qquad\qquad \doteq\,i\,(2\pi )^4\delta^{(4)}(p_1+p_2+p_l+p_{\nu}-p)\left (-i\,{\cal A}^{+-}\right )\,,
\end{eqnarray} 
with the on-shell conditions, 
\begin{equation}
p^2\,=\,M_{K^{\pm}}^2\,, \; p_1^2\,=\,M_{\pi^{\pm}}^2\,, \; p_2^2\,=\,M_{\pi^{\pm}}^2\,, \; p_l^2\,=\,m_l^2\,, \; p_{\nu}^2\,=\,0\,. 
\end{equation} 

\subsection{Form factors}

Let us introduce the notations, 
\begin{equation}
P\,=\,p_1+p_2\,, \;
Q\,=\,p_1-p_2\,, \; L\,=\,p_l+p_{\nu}\,, \;
N\,=\,p_l-p_{\nu}\,. 
\end{equation} 
By Lorentz covariance, the decay amplitude can be parameterized in terms of three vectorial form factors, $f^{+-}$, $g^{+-}$, $r^{+-}$, one anomalous form factor, $h^{+-}$, and one tensorial form factor, $T$,
\begin{eqnarray}
{\cal A}^{+-}
&\doteq & \frac{G_FV_{us}^*}{\sqrt{2}}\,{\overline u}(\boldsymbol{p}_{\nu})
        (1+\gamma^5)\left [\frac{1}{M_{K^{\pm}}}\,(f^{+-}P^{\mu}+g^{+-}Q^{\mu}
+r^{+-}L^{\mu})\gamma^{\mu}\right.
\nonumber \\
&+& \left. i\,\frac{h^{+-}}{M_{K^{\pm}}^3}\,\epsilon_{\mu\nu\rho\sigma}
        \gamma^{\mu}L^{\nu}P^{\rho}Q^{\sigma}
        -i\,\frac{T}{M_{K^{\pm}}^2}\,\sigma_{\mu\nu}\,p_1^{\mu}p_2^{\nu}\right
        ]v(\boldsymbol{p}_l)\,. 
\nonumber
\end{eqnarray}
In the preceding formula, $\sigma^{\mu\nu}$ is nothing else than the usual gamma matrix commutator,
$$
\sigma^{\mu\nu}\,\doteq\,\dfrac{i}{2}\left[ \gamma^{\mu},\gamma^{\nu}\right] \,,
$$ 
$V_{us}$ denotes the Cabibbo-Kobayashi-Maskawa flavor-mixing matrix element, and $G_F$ is the so-called Fermi coupling constant. Note that the normalization factors are written in powers of the charged kaon mass by pure convention. Let $F^{+-}$, $G^{+-}$, $R^{+-}$, and $H^{+-}$, denote respectively the form factors, $f^{+-}$, $g^{+-}$, $r^{+-}$, and $h^{+-}$ in the Isospin limit. At Tree-level, these quantities are given by the Current Algebra formulae~\cite{Callan:1966hu, Weinberg:1966, Weinberg:1966E, Berman:1968ss}, 
\begin{eqnarray}
F^{+-}
&=& G^{+-} 
\nonumber \\
&=& \frac{M_{K^{\pm}}}{\sqrt{2}F_0}\,, 
\nonumber \\
R^{+-}
&=& \frac{M_{K^{\pm}}}{2\sqrt{2}F_0}\left
  (1+\frac{s_{\pi}+t_{\pi}-u_{\pi}}{s_l-M_{K^{\pm}}^2}\right )\,, 
\nonumber \\
H^{+-}
&=& 0\,. \nonumber
\end{eqnarray} 
Herein, the Lorentz invariants, $s_{\pi}$ and $s_l$, represent the dipion and dilepton masses, respectively, 
\begin{equation}
s_{\pi}\,\doteq\,(p_1+p_2)^2\,, \qquad s_l\,\doteq\,(p_l+p_{\nu})^2\,.
\end{equation} 
The scalars, $t_{\pi}$ and $u_{\pi}$, denote exchange energies between $K^+$ and the two pions, $\pi^+$ and $\pi^-$, respectively,
\begin{equation}
t_{\pi}\,\doteq\,(p-p_1)^2\,, \qquad u_{\pi}\,\doteq\,(p-p_2)^2\,.
\end{equation}  

\subsection{Decay rate}

The decay rate $\Gamma$ for process (\ref{eq:process}) is obtained from the differential decay rate,
\begin{equation}
d\Gamma\,\doteq\,\frac{1}{2M_{K^{\pm}}}\,d\Phi\sum_{\mathrm{spins}}|{\cal A}^{+-}|^2\,,
\end{equation} 
by integrating over the differential phase space,
\begin{eqnarray}
d\Phi
&=& (2\pi )^4\delta^{(4)}(p_1+p_2+p_l+p_{\nu}-p)\times 
\nonumber \\
&& \frac{d^3\boldsymbol{p}_1}{(2\pi )^32E_1}\,
        \frac{d^3\boldsymbol{p}_2}{(2\pi )^32E_2}\,
        \frac{d^3\boldsymbol{p}_l}{(2\pi )^32E_l}\,
        \frac{d^3\boldsymbol{p}_{\nu}}{(2\pi
        )^32|\boldsymbol{p}_{\nu}|}\,,
\nonumber
\end{eqnarray}
with particle energies,
\begin{equation}
E_i\,=\,\sqrt{p_i^2+\boldsymbol{p}_i^2}\,.
\end{equation}
To this end, we will follow the approach used in~\cite{Cabibbo:1965} which consists on looking at $K_{\ell 4}$ decays as being two-body decays into a dipion of mass $s_{\pi}$ and a dilepton of mass $s_l$. The two systems subsequently
decay in their own center-of-mass frames. To describe the decay
distribution it is convenient to use, besides the invariant
masses, $s_{\pi}$ and $s_l$, the angles $\theta_{\pi}$,
$\theta_l$, and $\phi$ as illustrated in figure~\ref{fig:kinematics}. Then, one has to express all of the scalar products obtained from momenta, $p_1$, $p_2$,
$p_l$, and $p_{\nu}$, or equivalently from, $P$, $Q$, $L$, and
$N$, in terms of the five independent variables, $s_{\pi}$, $s_l$, $\theta_{\pi}$, $\theta_l$, and $\phi$. Naturally, Isospin breaking effects enter the expression of these scalar products through the difference between the two pion masses. Since the mesonic final state in process~(\ref{eq:process}) consists on two charged pions, one concludes that Isospin breaking does not affect scalar products,
\begin{eqnarray}
P^2
&=& s_{\pi}\,, 
\\
Q^2
&=& 4M_{\pi^{\pm}}^2-s_{\pi}\,, 
\\
L^2
&=& s_l\,, 
\\
N^2
&=& 2m_l^2-s_l\,, 
\\
P\cdot Q
&=& 0\,, 
\\
P\cdot L
&=& \frac{1}{2}\,(M_{K^{\pm}}^2-s_{\pi}-s_l)\,, 
\\
P\cdot N
&=& \frac{1}{2}\,(M_{K^{\pm}}^2-s_{\pi}-s_l)\,z_l+(1-z_l)X\cos\theta_l\,,
\\
Q\cdot L
&=& X\sigma_{\pi}\cos\theta_{\pi}\,,
\\
Q\cdot N
&=& z_lX\sigma_{\pi}\cos\theta_{\pi}
+\frac{1}{2}\,(1-z_l)\sigma_{\pi}\,[\,(M_{K^{\pm}}^2-s_{\pi}-s_l)\times
\nonumber \\ 
&& \cos\theta_{\pi}\cos\theta_l
-2\sqrt{s_{\pi}s_l}\sin\theta_{\pi}\sin\theta_l\cos\phi\,]\,,
\\
L\cdot N
&=& m_l^2\,, 
\\
\epsilon_{\mu\nu\rho\sigma}L^{\mu}N^{\nu}P^{\rho}Q^{\sigma}
&=& -(1-z_l)X\sigma_{\pi}\sqrt{s_{\pi}s_l}\sin\theta_{\pi}\sin\theta_l\sin\phi\,.
\end{eqnarray} 
Herein,
\begin{eqnarray}
z_l
&\doteq& \frac{m_l^2}{s_l}\,, 
\\
X
&\doteq& \frac{1}{2}\,\lambda^{1/2}(M_{K^{\pm}}^2,s_{\pi},s_l)\,, 
\\
\sigma_{\pi}
&\doteq& \sqrt{1-\dfrac{4M_{\pi^{\pm}}^2}{s_{\pi}}}\,,
\end{eqnarray}
and the function,
\begin{equation}
\lambda (x,y,z)\,\doteq\,x^2+y^2+z^2-2xy-2xz-2yz\,,
\end{equation}
is the usual K\"all\'en function.

The same holds for the differential phase space which reads in terms of the five independent variables, 
\begin{equation}
\label{eq:phase_space} 
d^5\Phi\,=\,M_{K^{\pm}}^3N(s_{\pi},s_l)ds_{\pi}ds_l
d(\cos\theta_{\pi})d(\cos\theta_l)d\phi\,,
\end{equation} 
with,
\begin{equation}
N(s_{\pi},s_l)\,\doteq\,\frac{1}{2^{13}\pi^6}\,
\frac{1}{M_{K^{\pm}}^5}\,(1-z_l)X\sigma_{\pi}\,.
\end{equation} 
Obviously, the phase space is reached by integrating (\ref{eq:phase_space}) over the following range of variables,
\begin{equation}
\begin{array}{ccccc}
0                & \leq & \phi         & \leq & 2\pi \\
0                & \leq & \theta_l     & \leq & \pi \\
0                & \leq & \theta_{\pi} & \leq & \pi \\
m_l^2            & \leq & s_l          & \leq & (M_{K^{\pm}}-\sqrt{s_{\pi}})^2 \\
4M_{\pi^{\pm}}^2 & \leq & s_{\pi}      & \leq & (M_{K^{\pm}}-m_l)^2\,.
\end{array}
\end{equation}

The last step consists on squaring the amplitude and summing over spins to get,
\begin{equation}
\sum_{\mathrm{spins}}|{\cal A}^{+-}|^2\,=\,2G_F^2|V_{us}|^2\,\frac{1}{M_{K^{\pm}}^2}\,
j_5(s_{\pi},s_l,\theta_{\pi},\theta_l,\phi )\,,
\end{equation} 
with the following expression for the intensity spectrum,
\begin{eqnarray}
j_5
&=& |f|^2\left[ (P\cdot L)^2-(P\cdot N)^2-s_{\pi}(s_l-m_l^2)\right] 
\nonumber \\ 
&+& |g|^2\left[ (Q\cdot L)^2-(Q\cdot N)^2-Q^2(s_l-m_l^2)\right] 
\nonumber \\ 
&+& |r|^2m_l^2(s_l-m_l^2) 
\nonumber \\ 
&-& |h|^2\,\dfrac{1}{M_{K^{\pm}}^4}\left\lbrace (\epsilon_{\mu\nu\rho\sigma}L^{\mu}N^{\nu}P^{\rho}Q^{\sigma})^2+(s_l-m_l^2)\left[ Q^2X^2+s_{\pi}(Q\cdot L)^2\right] \right\rbrace  
\nonumber \\ 
&+& (f^*g+fg^*)\left[ (P\cdot L)(Q\cdot L)-(P\cdot N)(Q\cdot N)\right] 
\nonumber \\ 
&+& (f^*r+fr^*)m_l^2\left[ (P\cdot L)-(P\cdot N)\right] 
\nonumber \\ 
&+& (f^*h+fh^*)\,\dfrac{1}{M_{K^{\pm}}^2}\left[ (Q\cdot N)(P\cdot L)^2\right.
\nonumber \\ 
&-& \left. (Q\cdot L)(P\cdot L)(P\cdot N)-s_{\pi}s_l(Q\cdot N)+m_l^2s_{\pi}(Q\cdot L)\right] 
\nonumber \\ 
&+& (g^*r+gr^*)m_l^2\left[ (Q\cdot L)-(Q\cdot N)\right] 
\nonumber \\ 
&+& (g^*h+gh^*)\,\dfrac{1}{M_{K^{\pm}}^2}\left[ (P\cdot L)(Q\cdot L)(Q\cdot N)\right. 
\nonumber \\ 
&-& \left. (P\cdot N)(Q\cdot L)^2+s_lQ^2(P\cdot N)-m_l^2Q^2(P\cdot L)\right] 
\nonumber \\ 
&-& \epsilon_{\mu\nu\rho\sigma}L^{\mu}N^{\nu}P^{\rho}Q^{\sigma}\,
\dfrac{i}{M_{K^{\pm}}^2}\left[ M_{K^{\pm}}^2(f^*g-fg^*)\right. 
\nonumber \\ 
&-& \left.(P\cdot N)(f^*h-fh^*)-(Q\cdot N)(g^*h-gh^*)-m_l^2(r^*h-rh^*)\right] 
\nonumber \\
&+& (fT^*+Tf^*)\,\dfrac{m_l}{2M_{K^{\pm}}}\left[ s_{\pi}(Q\cdot N)-s_{\pi}(Q\cdot L)\right] 
\nonumber \\ 
&+& (gT^*+Tg^*)\,\dfrac{m_l}{2M_{K^{\pm}}}\left[ Q^2(P\cdot L)-Q^2(P\cdot N)\right] 
\nonumber \\ 
&+& (rT^*+Tr^*)\,\dfrac{m_l}{2M_{K^{\pm}}}\left[ (P\cdot L)(Q\cdot N)-(P\cdot N)(Q\cdot L)\right] 
\nonumber \\ 
&+& (rT^*-Tr^*)\,\dfrac{im_l}{2M_{K^{\pm}}}\,\epsilon_{\mu\nu\rho\sigma}
L^{\mu}N^{\nu}P^{\rho}Q^{\sigma} 
\nonumber \\ 
&-& (hT^*+Th^*)\,\dfrac{m_l}{2M_{K^{\pm}}^3}\left[ -Q^2(P\cdot L)^2+Q^2(P\cdot L)(P\cdot N)-s_{\pi}(Q\cdot L)^2\right.
\nonumber \\ 
&+& \left. s_{\pi}(Q\cdot L)(Q\cdot N)+s_{\pi}s_lQ^2-s_{\pi}Q^2(L\cdot N)\right] 
\nonumber \\ 
&+& |T|^2\,\dfrac{1}{8M_{K^{\pm}}^2}\left[ s_{\pi}s_lQ^2-2Q^2(P\cdot L)^2-2s_{\pi}(Q\cdot L)^2\right.
\nonumber \\ 
&-& \left. s_{\pi}Q^2N^2+2Q^2(P\cdot N)^2+2s_{\pi}(Q\cdot N)^2\right] \,.
\end{eqnarray} 
From the foregoing we conclude that, apart from the corrections affecting the form factors $F^{+-}$, $G^{+-}$, $R^{+-}$ and $H^{+-}$ directly, Isospin breaking enters the expression of the intensity spectrum through the additional terms proportional to the tensorial form factor, $T$.   

%% file: section_2.tex
\section{Form factors to one-loop level}

We present here a one-loop calculation of the $K_{\ell 4}$ decay amplitude for the process (\ref{eq:process}) including isospin breaking terms. The starting point is an effective chiral Lagrangian describing the dynamics of mesons, photons and leptons.

\subsection{The effective Lagrangian}

In order to treat completely electromagnetic effects in $K_{\ell 4}$
decays, not only the pseudoscalars but also the photon
and the light leptons have to be included as dynamical degrees of freedom
in an appropriate effective Lagrangian~\cite{Knecht:1999ag}. The starting point is QCD in the limit $m_u = m_d = m_s = 0$. The resulting chiral symmetry, $G = SU(3)_L \times SU(3)_R$, is spontaneously broken to $SU(3)_V$. The pseudoscalar mesons $(\pi ,K,\eta )$ are nothing else than the corresponding Goldstone fields $\phi_i$ ($i = 1,\ldots ,8$) acting as coordinates of the coset space $SU(3)_L \times SU(3)_R/SU(3)_V$.
The transformation rules for the coset variables $u_{L,R}(\phi )$ are
\begin{eqnarray}
u_L(\phi )
&\stackrel{G}{\rightarrow}& g_L u_L h(g,\phi )^{-1}\,,
\nonumber \\
u_R(\phi)
&\stackrel{G}{\rightarrow}& g_R u_R h(g,\phi)^{-1}\,,
\nonumber \\
g = (g_L,g_R)
&\in & SU(3)_L \times SU(3)_R\,,
\end{eqnarray}
where $h(g,\phi )$ is the nonlinear realization of $G$~\cite{Coleman:1969sm, Callan:1969sn}.

As stated before, the photon field $A_{\mu}$ and the leptons $\ell ,\nu_{\ell}$ ($\ell =\mathrm{e},\mu$) have to be dynamical. Thus, they most be introduced in
the covariant derivative,
\begin{equation}
u_{\mu}\,\doteq\,i\,[\,u_R^{\dagger}(\partial_{\mu}-ir_{\mu})u_R-u_L^{\dagger}
(\partial_{\mu}-il_{\mu})u_L\,]\,,
\end{equation}
by adding appropriate terms to the usual external vector and axial-vector
sources ${\cal V}_{\mu}$, ${\cal A}_{\mu}$. At the quark level, this procedure corresponds to the usual minimal coupling prescription in the case of electromagnetism, and to Cabibbo universality in the case of the charged weak currents,
\begin{eqnarray}
\label{sources}
l_{\mu}
&\doteq & v_{\mu}-a_{\mu}-eQ_L^{\mathrm{em}}A_{\mu}+\sum_{\ell}
(\overline{\ell}\gamma_{\mu}\nu_{\ell L}Q_L^{\mathrm{w}}+\overline{\nu_{\ell L}}
\gamma_{\mu} \ell Q_L^{{\mathrm{w}}\dagger})\,,
\nonumber \\
r_{\mu}
&\doteq & v_{\mu}+a_{\mu}-eQ_R^{\mathrm{em}}A_{\mu}\,.
\end{eqnarray}
The $3 \times 3$ matrices $Q_{L,R}^{\mathrm{em}}$, $Q_L^{\mathrm{w}}$ are
spurion fields corresponding to electromagnetic and weak coupling, respectively. They transform as,
\begin{equation}
Q_L^{\mathrm{em,w}} \stackrel{G}{\rightarrow} g_L Q_L^{\mathrm{em,w}} g_L^{\dagger}\,, \qquad
Q_R^{\mathrm{em}} \stackrel{G}{\rightarrow} g_R Q_R^{\mathrm{em}} g_R^{\dagger}\,,
\end{equation}
under the chiral group. In practical calculations, one identifies $Q_{L,R}^{\mathrm{em}}$ with the quark charge matrix
\begin{equation}
\label{Qem}
Q^{\mathrm{em}}\,\doteq\,\left ( \begin{array}{ccc}
                            2/3 & 0    & 0 \\
                              0 & -1/3 & 0 \\
                              0 & 0    & -1/3
                            \end{array}\right )\,,
\end{equation}
whereas the weak spurion is replaced by,
\begin{equation}
\label{Qw}
Q_L^{\mathrm{w}}\,\doteq\, -2\sqrt{2}\; G_F \left ( \begin{array}{ccc}
                                               0 & V_{ud} & V_{us} \\
                                               0 & 0      & 0 \\
                                               0 & 0      & 0
                                               \end{array} \right )\,,
\end{equation}
where $G_F$ is the Fermi coupling constant and $V_{ud}$, $V_{us}$ are
Kobayashi-Maskawa matrix elements.

In order to take into account electromagnetic mass difference between charged and neutral mesons, it is convenient to define the following electromagnetic and weak sources,
\begin{equation}
\label{Qhom}
{\cal Q}_L^{\mathrm{em,w}}\,\doteq\,u_L^{\dagger} Q_L^{\mathrm{em,w}} u_L, \qquad
{\cal Q}_R^{\mathrm{em}}\,\doteq\,u_R^{\dagger} Q_R^{\mathrm{em}} u_R\,
\end{equation}
transforming as
\begin{eqnarray}
{\cal Q}_L^{\mathrm{em,w}}
&\stackrel{G}{\rightarrow}& h(g,\phi ) {\cal Q}_L^{\mathrm{em,w}} h(g,\phi )^{-1}\,,
\nonumber \\
{\cal Q}_R^{\mathrm{em}}
&\stackrel{G}{\rightarrow}& h(g,\phi ) {\cal Q}_R^{\mathrm{em}} h(g,\phi )^{-1}\,.
\end{eqnarray}

With these building blocks, the lowest order effective Lagrangian takes
the form
\begin{eqnarray}
\label{Leff}
{\cal L}_{\mathrm{eff}}
&=& \frac{{F_0}^2}{4} \; \langle u_{\mu} u^{\mu} + \chi_+\rangle +
e^2 {F_0}^4 Z_0 \langle {\cal Q}_L^{\mathrm{em}} {\cal Q}_R^{\mathrm{em}}\rangle \nonumber \\
&-& \frac{1}{4} F_{\mu\nu} F^{\mu\nu} + \sum_{\ell}
[\overline{\ell}(i \! \not\!\partial + e \! \not\!\!A - m_l)\ell +
\overline{\nu}_{\ell L} \, i \! \not\!\partial \nu_{\ell L}],
\end{eqnarray}
where $\langle \;\rangle$ denotes the trace in three-dimensional
flavor space.

The low-energy constant $F_0$ appearing in the preceding formula is an order parameter for chiral symmetry since it testifies to its spontaneous breaking. It represents the pion decay constant in the chiral limit, $m_u=m_d=m_s=0$, and in the absence of electroweak interactions. Explicit chiral symmetry breaking due to quark masses is included in,
$$
\chi_+\,\doteq\,u_R^{\dagger} \chi u_L + u_L^{\dagger} \chi^{\dagger} u_R\,.
$$
In practice, one makes the following substitution,
\begin{equation}
\chi\,\rightarrow\,2{B_0} {\cal M}_{\mathrm{quark}}\,\doteq\,2B_0\left(
\begin{array}{ccc}
m_u & 0 & 0 \\
0   & m_d & 0 \\
0   &     & m_s
\end{array}\right)\,,
\end{equation}
where $B_0$ is an order parameter for chiral symmetry. It is related to the quark condensate in the chiral limit by,
\begin{equation}
\langle\overline{q}q\rangle\,=\,-F_0^2B_0\,.
\end{equation}
The low-energy constant $Z_0$ expresses explicit chiral symmetry breaking by electromagnetism. It is given by the electromagnetic mass of the pion as we will see below.

In the absence of electroweak interactions, that is, for $m_u=m_d\,, \; \alpha=0$, ChPT is Isospin-invariant. In order to study Isospin breaking effects in ChPT processes, the usual chiral expansion in powers of $p$ and $m_q$ is no more sufficient. One must also expand matrix elements in powers of the isospin breaking parameters, $m_d-m_u$ and $\alpha$. On the other hand, the best accuracy ever reached in strong interaction observable measurements does not exceed the $5\%$ level. Thus, we consider that an expansion to orders ${\cal O}(m_d-m_u)$ and ${\cal O}(\alpha )$ is highly adequate for our purposes. Moreover, chiral expansion and Isospin breaking expansion have to be related in a consistent way in order to obtain reliable results. We adopt an expansion scheme where the Isospin breaking parameters are considered as quantities of order $p^2$ in the chiral counting,
\begin{equation}
\label{eq:leading_order}
{\cal O}(m_d-m_u)\,=\,{\cal O}(\alpha )\,=\,{\cal O}(m_q)\,=\,{\cal O}(p^2)\,.
\end{equation}
Therefore, tree level calculation corresponding to leading chiral
order is characterized by chiral orders cited in
(\ref{eq:leading_order}). Concerning one-loop level calculation
which corresponds to next-to-leading chiral order, it is
characterized by the following chiral orders,
$$
\mathcal{O}(p^4)\,,\;\mathcal{O}(m_q^2)\,,\;\mathcal{O}(p^2m_q)\,,
$$
\begin{equation}
\label{eq:next_to_leading_order}
\mathcal{O}\left( p^2(m_d-m_u)\right)\,,\;\mathcal{O}\left( m_q(m_d-m_u)\right)\,,\;\mathcal{O}(p^2\alpha )\,,\;\mathcal{O}(m_q\alpha )\,.
\end{equation}

We have now all necessary elements to calculate any Green function in the framework of ChPT including Isospin breaking effects. For example, the leading chiral order expressions for light meson masses are found to be,
\begin{eqnarray}
M_{\pi^0}^2
&=& M_{\pi}^2\,,
\label{eq:neutral_pion} \\
M_{\pi^{\pm}}^2
&=& M_{\pi}^2+2Z_0e^2F_0^2\,,
\label{eq:charged_pion} \\
M_{K^0}^2
&=& M_K^2+\dfrac{2\epsilon}{\sqrt{3}}\left( M_K^2-M_{\pi}^2\right) \,,
\label{eq:neutral_kaon} \\
M_{K^{\pm}}^2
&=& M_K^2-\dfrac{2\epsilon}{\sqrt{3}}\left( M_K^2-M_{\pi}^2\right) +2Z_0e^2F_0^2\,,
\label{eq:charged_kaon} \\
M_{\eta}^2
&=& \dfrac{1}{3}\left( 4M_K^2-M_{\pi}^2\right) \,.
\label{eq:eta_mass}
\end{eqnarray}
Herein, $M_{\pi}$ and $M_K$ represent respectively pion and kaon masses in the absence of isospin breaking,
\begin{equation}
\label{eq:quark_masses}
M_{\pi}^2\,\doteq\,2B_0\hat{m}\,, \quad M_K^2\,\doteq\,B_0(\hat{m}+m_s)\,, \quad 2\hat{m}\,\doteq\,m_u+m_d\,,
\end{equation}
$\epsilon$ measures the rate of $SU(2)$ to $SU(3)$ breaking,
\begin{equation}
\label{eq:epsilon}
\epsilon\,\doteq\,\dfrac{\sqrt{3}}{4}\,\dfrac{m_d-m_u}{m_s-\hat{m}}\,.
\end{equation}
At next-to-leading chiral order, one-loop calculation is involved.
Vertices are extracted from Lagrangian (\ref{Leff}). Meson masses
in the propagators as well as in the vertices can be identified
with expressions (\ref{eq:neutral_pion})-(\ref{eq:eta_mass}). As
is well known, loops are ultraviolet divergent. To remove
divergences, renormalization should be employed. The procedure
consists on adding to Lagrangian (\ref{Leff}) suitable
counter-terms~\cite{Gasser:1985gg, Urech:1995hd, Neufeld:1995eg,
Neufeld:1996mu, Knecht:1999ag} generating exactly the same
divergences but with opposite sign. Moreover, the cancellation
should occur \textit{order by order} in the chiral expansion as
dictated by renormalizability principles of effective field
theories. Counter-terms are modulated by low-energy constants
which are order parameters for chiral symmetry. In order to
determine these constants one proceeds as follows. Let $C$ be
either a low-energy constant or a combination of low-energy
constants and $\Upsilon$ an observable very sensitive to
variations of $C$. One first calculate the expression of
$\Upsilon$ in the framework of ChPT to any given order and then
\textit{match} the obtained expression with an experimental
measurement of $\Upsilon$. It is clear that the value for $C$
deduced from this matching does not constitute a \textit{genuine
determination} of the low-energy constant. In fact, it represents
the value of $C$ at the given chiral order and with the accuracy
of the experimental measurement. Note that the method of effective
Lagrangian has the disadvantage of an infinitely increasing number
of low-energy constants when going to higher and higher orders in
the low-energy expansion. For instance, two constants in the
strong sector, $B_0$ and $F_0$, and one constant in the
electroweak meson sector, $Z_0$, parameterize the leading chiral
order. At next-to-leading chiral order, one has ten low-energy
constants in the strong sector, $L_1\,, \ldots\,, L_{10}$,
fourteen constants in the electroweak meson sector, $K_1\,,
\ldots\,, K_{14}$, and seven constants in the electroweak leptonic
sector, $X_1\,, \ldots\,, X_7$. The constants, $L_i$, $K_i$ and
$X_i$, are divergent. They absorb the divergence of loops via the
renormalization,
\begin{eqnarray}
L_i
&\doteq & L_i^r(\mu )+\Gamma_i\,\overline{\lambda}\,, \qquad i\,=\,1\,,\ldots\,, 10\,,
\\
K_i
&\doteq & K_i^r(\mu )+\Sigma_i\,\overline{\lambda}\,, \qquad i\,=\,1\,,\ldots\,, 14\,,
\\
X_i
&\doteq & X_i^r(\mu )+\Xi_i\,\overline{\lambda}\,, \qquad i\,=\,1\,,\ldots\,, 7\,.
\end{eqnarray}
Herein, $\overline{\lambda}$ corresponds to pole subtraction in the $\overline{\mathrm{MS}}$ dimensional regularization scheme (see appendix). The beta-functions, $\Gamma_i$, $\Sigma_i$ and $\Xi_i$, can be found in~\cite{Gasser:1985gg}, \cite{Urech:1995hd} and~\cite{Knecht:1999ag}, respectively. The scale $\mu$ cancels in observables as can be seen from the renormalization group equations,
\begin{equation}
L_i^r(\mu_2)\,=\,L_i^r(\mu_1)+\dfrac{\Gamma_i}{16\pi^2}\,\ln\dfrac{\mu_1}{\mu_2}\,,
\end{equation}
and similar for $K_i$ and $X_i$.

\subsection{Leading order}

Let us start with the tree-level calculation corresponding to the leading chiral order. The different topologies for $K_{\ell 4}$ decays at tree level in
perturbation theory are drawn in figure~\ref{fig:tree}. Decay amplitudes for the different processes were calculated in~\cite{Nehme:2003bz} where the expressions for the corrected form factors in the charged channel were found to be,
\begin{eqnarray}
f^{+-}
&=& F^{+-}\,,
\nonumber \\
g^{+-}
&=& G^{+-}\,,
\nonumber \\
r^{+-}
&=& R^{+-}+4Z_0e^2F_0^2\,\frac{M_{K^{\pm}}}{\sqrt{2}F_0}\,\frac{1}{s_l-M_{K^{\pm}}^2} \nonumber \\
&-& \frac{2e^2F_0^2}{s_{\pi}}\left
  (\frac{t_{\pi}-u_{\pi}}{s_l-M_{K^{\pm}}^2}-
        \frac{Q\cdot L+Q\cdot N}{M_{K^{\pm}}^2-m_l^2-2p\cdot p_{\nu}}\right )\frac{M_{K^{\pm}}}{\sqrt{2}F_0}\,,
\nonumber \\
T
&=& \frac{4e^2F_0^2}{s_{\pi}}\,\frac{m_lM_{K^{\pm}}}{M_{K^{\pm}}^2-m_l^2-2p\cdot p_{\nu}}\,\frac{M_{K^{\pm}}}{\sqrt{2}F_0}\,.
\nonumber
\end{eqnarray}

\subsection{Born contribution}

We proceed with the calculation of form factors $f^{+-}$ and $g^{+-}$ to next-to-leading chiral order. Feynman diagrams representing the amplitude will be separated into two sets: photonic and non photonic diagrams. The non photonic set is drawn in figure~\ref{fig:strong}. The calculation of these diagrams is standard in field theory. The starting point is Lagrangian (\ref{Leff}). For the non linear realization of chiral symmetry we will use the exponential parametrization,
\begin{equation}
\label{eq:exponential_parametrization}
u_R\,=\,u_L^{\dagger}\,=\,\mathrm{exp}\left\lbrace \dfrac{i\Phi}{2F_0}\right\rbrace \,,
\end{equation}
where $\Phi$ is the linear realization of $SU(3)$ and can be decomposed in the basis of Gellmann-Low matrices $\lambda_a$ as,
$$
\Phi\,=\,\Phi^{\dagger}\,=\,\sum_{a=1}^8\lambda_a\phi^a\,.
$$
In terms of physical fields the matrix $\Phi$ can be written,
\begin{eqnarray}
\Phi_{11}
&=& \left( 1+\dfrac{\tilde{\epsilon}_2}{\sqrt{3}}\right) \pi^0+\left( -\tilde{\epsilon}_1+\dfrac{1}{\sqrt{3}}\right) \eta\,,
\nonumber \\
\Phi_{12}
&=& -\sqrt{2}\pi^+\,, \; \Phi_{13}\,=\,-\sqrt{2}K^+\,, \; \Phi_{21}\,=\,\sqrt{2}\pi^-\,,
\nonumber \\
\Phi_{22}
&=& \left( -1+\dfrac{\tilde{\epsilon}_2}{\sqrt{3}}\right) \pi^0+\left( \tilde{\epsilon}_1+\dfrac{1}{\sqrt{3}}\right) \eta\,,
\nonumber \\
\Phi_{23}
&=& -\sqrt{2}K^0\,, \; \Phi_{31}\,=\,\sqrt{2}K^-\,, \; \Phi_{32}\,=\,-\sqrt{2}\overline{K}^0\,,
\nonumber \\
\Phi_{33}
&=& -\dfrac{2}{\sqrt{3}}\left(\tilde{\epsilon}_2\pi^0+\eta\right) \,.
\label{eq:linear_parametrization}
\end{eqnarray}
The two mixing angles $\tilde{\epsilon}_1$ and $\tilde{\epsilon}_2$ relate $\phi^3$ and $\phi^8$ to the mass eigenstates $\pi^0$ and $\eta$,
\begin{equation}
\left(
\begin{array}{c}
\pi^0 \\
\eta
\end{array}
\right) \,\doteq\,
\left(
\begin{array}{cc}
1                   & \tilde{\epsilon}_1 \\
-\tilde{\epsilon}_2 & 1
\end{array}
\right)
\left(
\begin{array}{c}
\phi^3 \\
\phi^8
\end{array}
\right) \,.
\end{equation}
Notice that $\tilde{\epsilon}_1=\tilde{\epsilon}_2=\epsilon$ at leading order. The next step consists on expanding Lagrangian (\ref{Leff}) to fifth order in pseudoscalar fields, generating Feynman rules and drawing allowed topologies. We then calculate pseudoscalar propagators and derive masses and wave function renormalization constants. Finally, we expand the next-to-leading order Lagrangian to third order in pseudoscalar fields and obtain the counterterm contribution.

Let us denote by $\delta F$ and $\delta G$ the next-to-leading
order corrections to the $F^{+-}$ and $G^{+-}$ form factors,
respectively,
\begin{eqnarray}
f^{+-}
  &=& \frac{M_{K^{\pm}}}{\sqrt{2}F_0}\,\bigg (\,1+\delta F\,\bigg )\,,
  \nonumber \\
g^{+-}
  &=& \frac{M_{K^{\pm}}}{\sqrt{2}F_0}\,\bigg (\,1+\delta G\,\bigg )\,.
  \nonumber
\end{eqnarray}
The expressions for $\delta F$ and $\delta G$ are lengthy. Therefore we will separate them to different contributions depending on the topology of the Feynman diagram representing a given contribution. Notice that in the following we will use the same notations as in reference~\cite{Nehme:2003bz} where definitions and expressions for loop integrals have been given in the appendix.  

Let us start with Born contribution. This contribution is obtained from diagram (a) in figure~\ref{fig:strong}. We take the corresponding vertex from Lagrangian (\ref{Leff}) and multiply by the wave function renormalization constant factor to obtain,
\begin{eqnarray}
\delta F
&=& \delta G
\nonumber \\
&=& -\dfrac{1}{24F_0^2}\left[ 3\left( 3+\dfrac{2\epsilon}{\sqrt{3}}\right) A_0(M_{\pi^0})+3\left( 1-\dfrac{2\epsilon}{\sqrt{3}}\right) A_0(M_{\eta})\right.
\nonumber \\
&& \left.+10A_0(M_{\pi^{\pm}})+6A_0(M_{K^0})+8A_0(M_{K^{\pm}})\right]
\nonumber \\
&-& \dfrac{4}{F_0^2}\left[ 3(M_{\pi}^2+2M_K^2)L_4+(M_{K^{\pm}}^2-M_{\pi^{\pm}}^2+3M_{\pi^0}^2)L_5\right]
\nonumber \\
&+& \dfrac{e^2}{2}\left[ \dfrac{2}{M_K^2}\,A_0(M_K)+\dfrac{4}{M_{\pi}^2}\,A_0(M_{\pi})-\dfrac{1}{m_l^2}\,A_0(m_l)\right.
\nonumber \\
&-& \left.\dfrac{1}{16\pi^2}\left( 9
+2\ln\dfrac{m_{\gamma}^2}{M_K^2}
+4\ln\dfrac{m_{\gamma}^2}{M_{\pi}^2}
+2\ln\dfrac{m_{\gamma}^2}{m_l^2}\right) \right]
\nonumber \\
&-& \dfrac{e^2}{6}\left( 24K_1+24K_2+20K_5+20K_6+3X_6\right) \,.
\end{eqnarray}
The Born contribution is infrared divergent. This divergence emerges from the wave function renormalization constants of charged kaon, pions, and lepton.

\subsection{Counter-terms contribution}

The contribution in question follows from diagram (a) in figure~\ref{fig:strong} with the help of the next-to-leading order Lagrangian and reads,
\begin{eqnarray}
\delta F
&=& \dfrac{2}{F_0^2}\left[ 32p_1\cdot p_2L_1+4(M_{K^{\pm}}^2+2M_{\pi^{\pm}}^2-m_l^2+2p_1\cdot p_2-2p_l\cdot p_{\nu})L_2\right.
\nonumber \\
&+& (M_{K^{\pm}}^2+2M_{\pi^{\pm}}^2-m_l^2+10p_1\cdot p_2+2p\cdot p_1-2p\cdot p_2-2p_l\cdot p_{\nu})L_3
\nonumber \\
&+& 4(2M_K^2+5M_{\pi}^2)L_4+2(M_{K^{\pm}}^2-M_{\pi^{\pm}}^2
+3M_{\pi^0}^2)L_5
\nonumber \\
&& \qquad \left.+(m_l^2+2p_l\cdot p_{\nu})L_9\right]
\nonumber \\
&+& \dfrac{2e^2}{9}\left( 12K_1+84K_2+19K_5+37K_6+9K_{12}-30X_1\right) \,,
\nonumber \\
\delta G
&=& \dfrac{2}{F_0^2}\left[ -8(p\cdot p_1-p\cdot p_2)L_2\right.
\nonumber \\
&-& (M_{K^{\pm}}^2+2M_{\pi^{\pm}}^2-m_l^2+2p_1\cdot p_2+2p\cdot p_1-2p\cdot p_2-2p_l\cdot p_{\nu})L_3
\nonumber \\
&+& \left.4(M_{\pi}^2+2M_K^2)L_4
+2(M_{K^{\pm}}^2-M_{\pi^{\pm}}^2+3M_{\pi^0}^2)L_5
+(m_l^2+2p_l\cdot p_{\nu})L_9\right]
\nonumber \\
&+& \dfrac{2e^2}{9}\left( 12K_1+12K_2\right.
\nonumber \\
&& \quad \left.+36K_3+18K_4+7K_5+25K_6+9K_{12}+6X_1\right)\,.
\end{eqnarray}

\subsection{Tadpole contribution}

Tadpoles are shown in diagram (c) of figure~\ref{fig:strong} and
contribute to the form factors by the following,
\begin{eqnarray}
\delta F
&=& \dfrac{1}{12F_0^2}\left[ A_0(M_{\pi^0})+9A_0(M_{\eta})\right.
\nonumber \\
&& \left.+8A_0(M_{K^0})+4A_0(M_{\pi^{\pm}})+8A_0(M_{K^{\pm}})\right] \,,
\nonumber \\
\delta G
  &=& \dfrac{1}{4F_0^2}\left[ A_0(M_{\pi^0})+A_0(M_{\eta})+4A_0(M_{\pi^{\pm}})+4A_0(M_{K^{\pm}})\right]\,.
\end{eqnarray}

\subsection{The $s$-channel contribution}

The remaining contribution to the non photonic part of the decay amplitude comes from loop diagrams with two pseudoscalar propagators. This two-point function contribution will be separated in three parts depending on the Lorentz scalar governing its underlying kinematics. The $s$-channel contribution comes from diagram (c) in figure~\ref{fig:strong} and reads,
\begin{eqnarray}
\delta F
&=& \dfrac{1}{6F_0^2}\left\lbrace -2\left( 1+\dfrac{3\epsilon}{\sqrt{3}}\right) A_0(M_{\pi^0})\right.
\nonumber \\
&& +\dfrac{6\epsilon}{\sqrt{3}}\,A_0(M_{\eta})
-4A_0(M_{\pi^{\pm}})-2A_0(M_{K^0})-4A_0(M_{K^{\pm}})
\nonumber \\
&+& 3\left[ 2M_{\pi^{\pm}}^2-M_{\pi^0}^2+2p_1\cdot p_2\right.
\nonumber \\
&& \quad \left.+\dfrac{6\epsilon}{\sqrt{3}}\,(M_{\pi}^2+2p_1\cdot p_2)\right] B_0(-p_1-p_2,M_{\pi^0},M_{\pi^0})
\nonumber \\
&+& 3\left( 1-\dfrac{2\epsilon}{\sqrt{3}}\right) M_{\pi^0}^2B_0(-p_1-p_2,M_{\eta},M_{\eta})
\nonumber \\
&-& \dfrac{12\epsilon}{\sqrt{3}}\,(M_{\pi}^2+3p_1\cdot p_2)B_0(-p_1-p_2,M_{\pi},M_{\eta})
\nonumber \\
&-& 8(4M_{\pi^{\pm}}^2-3M_{\pi^0}^2+p_1\cdot p_2)B_1(-p_1-p_2,M_{\pi^{\pm}},M_{\pi^{\pm}})
\nonumber \\
&-& 4(M_{\pi^{\pm}}^2+p_1\cdot p_2)B_1(-p_1-p_2,M_{K^0},M_{K^0})
\nonumber \\
&-& 8(4M_{\pi^{\pm}}^2-3M_{\pi^0}^2+p_1\cdot p_2)B_1(-p_1-p_2,M_{K^{\pm}},M_{K^{\pm}})
\nonumber \\
&+& 4B_{00}(-p_1-p_2,M_{\pi^{\pm}},M_{\pi^{\pm}})
\nonumber \\
&+& 2B_{00}(-p_1-p_2,M_{K^0},M_{K^0})
\nonumber \\
&+& 4B_{00}(-p_1-p_2,M_{K^{\pm}},M_{K^{\pm}})
\nonumber \\
&+& 8(M_{\pi^{\pm}}^2+p_1\cdot p_2)B_{11}(-p_1-p_2,M_{\pi^{\pm}},M_{\pi^{\pm}})
\nonumber \\
&+& 4(M_{\pi^{\pm}}^2+p_1\cdot p_2)B_{11}(-p_1-p_2,M_{K^0},M_{K^0})
\nonumber \\
&+& \left.8(M_{\pi^{\pm}}^2+p_1\cdot p_2)B_{11}(-p_1-p_2,M_{K^{\pm}},M_{K^{\pm}})\right\rbrace \,,
\nonumber \\
\delta G
&=& -\dfrac{1}{F_0^2}\left\lbrace 2B_{00}(-p_1-p_2,M_{\pi^{\pm}},M_{\pi^{\pm}})\right.
\nonumber \\
&& \left.-B_{00}(-p_1-p_2,M_{K^0},M_{K^0})
+2B_{00}(-p_1-p_2,M_{K^{\pm}},M_{K^{\pm}})\right\rbrace \,.
\nonumber
\end{eqnarray}

\subsection{The $t$-channel contribution}

Diagram (d) in figure~\ref{fig:strong} generates the somewhat lengthy $t$-channel contribution,
\begin{eqnarray}
\delta F
&=& -\dfrac{1}{12F_0^2}\left\lbrace -6A_0(M_{K^0})-2A_0(M_{K^{\pm}})\right.
\nonumber \\
&+& 6\left[ M_{\pi^{\pm}}^2-M_{\pi^0}^2-p\cdot p_1\right.
\nonumber \\
&& \left.+\dfrac{2\epsilon}{\sqrt{3}}\,(M_K^2-M_{\pi}^2-2p\cdot p_1)\right] B_0(p_1-p,M_{\pi^0},M_{K^0})
\nonumber \\
&-& 3\left[ 3M_{\pi^0}^2-M_{\eta}^2-2M_{\pi^{\pm}}^2+2p\cdot p_1\right.
\nonumber \\
&& \left.-\dfrac{\epsilon}{\sqrt{3}}\,(M_{\pi}^2-M_{\eta}^2+8p\cdot p_1)\right] B_0(p_1-p,M_{\eta},M_{K^0})
\nonumber \\
&-& 6\left[ M_{K^{\pm}}^2+M_{\pi^{\pm}}^2-M_{\pi^0}^2-2p\cdot p_1\right.
\nonumber \\
&& \left.+\dfrac{2\epsilon}{\sqrt{3}}\,(2M_K^2-p\cdot p_1)\right] B_1(p_1-p,M_{\pi^0},M_{K^0})
\nonumber \\
&+& 3\left[ 2M_{K^{\pm}}^2-2M_{\pi^{\pm}}^2-3M_{\pi^0}^2+M_{\eta}^2\right.
\nonumber \\
&& \left.+\dfrac{4\epsilon}{\sqrt{3}}\,(M_{\pi}^2+M_{\eta}^2-p\cdot p_1)\right] B_1(p_1-p,M_{\eta},M_{K^0})
\nonumber \\
&-& 4(4M_{\pi^{\pm}}^2-3M_{\pi^0}^2-p\cdot p_1)B_1(p_1-p,M_{\pi^{\pm}},M_{K^{\pm}})
\nonumber \\
&+& 12\left( 1-\dfrac{5\epsilon}{\sqrt{3}}\right) B_{00}(p_1-p,M_{\pi^0},M_{K^0})
\nonumber \\
&+& 12\left( 2+\dfrac{5\epsilon}{\sqrt{3}}\right) B_{00}(p_1-p,M_{\eta},M_{K^0})
\nonumber \\
&+& 20B_{00}(p_1-p,M_{\pi^{\pm}},M_{K^{\pm}})
\nonumber \\
&+& 6\left[ M_{K^{\pm}}^2-p\cdot p_1\right.
\nonumber \\
&& \left.-\dfrac{2\epsilon}{\sqrt{3}}\,(2M_K^2-M_{\pi}^2-p\cdot p_1)\right] B_{11}(p_1-p,M_{\pi^0},M_{K^0})
\nonumber \\
&+& 6\left[ M_{K^{\pm}}^2-2M_{\pi^{\pm}}^2+p\cdot p_1\right.
\nonumber \\
&& \left.+\dfrac{2\epsilon}{\sqrt{3}}\,(2M_K^2-M_{\pi}^2-p\cdot p_1)\right] B_{11}(p_1-p,M_{\eta},M_{K^0})
\nonumber \\
&+& \left.4(2M_{K^{\pm}}^2-M_{\pi^{\pm}}^2-p\cdot p_1)B_{11}(p_1-p,M_{\pi^{\pm}},M_{K^{\pm}})\right\rbrace \,,
\nonumber \\
\delta G
&=& \dfrac{1}{12F_0^2}\left\lbrace -6A_0(M_{K^0})-2A_0(M_{K^{\pm}})\right.
\nonumber \\
&+& 6\left[ M_{\pi^{\pm}}^2-M_{\pi^0}^2-p\cdot p_1\right.
\nonumber \\
&& \left.+\dfrac{2\epsilon}{\sqrt{3}}\,(M_K^2-M_{\pi}^2-2p\cdot p_1)\right] B_0(p_1-p,M_{\pi^0},M_{K^0})
\nonumber \\
&-& 3\left[ 3M_{\pi^0}^2-M_{\eta}^2-2M_{\pi^{\pm}}^2+2p\cdot p_1\right.
\nonumber \\
&& \left.-\dfrac{\epsilon}{\sqrt{3}}\,(M_{\pi}^2-M_{\eta}^2+8p\cdot p_1)\right] B_0(p_1-p,M_{\eta},M_{K^0})
\nonumber \\
&-& 6\left[ M_{K^{\pm}}^2+M_{\pi^{\pm}}^2-M_{\pi^0}^2-2p\cdot p_1\right.
\nonumber \\
&& \left.+\dfrac{2\epsilon}{\sqrt{3}}\,(2M_K^2-p\cdot p_1)\right] B_1(p_1-p,M_{\pi^0},M_{K^0})
\nonumber \\
&+& 3\left[ 2M_{K^{\pm}}^2-2M_{\pi^{\pm}}^2-3M_{\pi^0}^2+M_{\eta}^2\right.
\nonumber \\
&& \left.+\dfrac{4\epsilon}{\sqrt{3}}\,(M_{\pi}^2+M_{\eta}^2-p\cdot p_1)\right] B_1(p_1-p,M_{\eta},M_{K^0})
\nonumber \\
&-& 4(4M_{\pi^{\pm}}^2-3M_{\pi^0}^2-p\cdot p_1)B_1(p_1-p,M_{\pi^{\pm}},M_{K^{\pm}})
\nonumber \\
&+& \dfrac{12\epsilon}{\sqrt{3}}\,B_{00}(p_1-p,M_{\pi},M_K)
\nonumber \\
&-& 12\left( 1+\dfrac{\epsilon}{\sqrt{3}}\right) B_{00}(p_1-p,M_{\eta},M_{K^0})
\nonumber \\
&-& 4B_{00}(p_1-p,M_{\pi^{\pm}},M_{K^{\pm}})
\nonumber \\
&+& 6\left[ M_{K^{\pm}}^2-p\cdot p_1\right.
\nonumber \\
&& \left.-\dfrac{2\epsilon}{\sqrt{3}}\,(2M_K^2-M_{\pi}^2-p\cdot p_1)\right] B_{11}(p_1-p,M_{\pi^0},M_{K^0})
\nonumber \\
&+& 6\left[ M_{K^{\pm}}^2-2M_{\pi^{\pm}}^2+p\cdot p_1\right.
\nonumber \\
&& \left.+\dfrac{2\epsilon}{\sqrt{3}}\,(2M_K^2-M_{\pi}^2-p\cdot p_1)\right] B_{11}(p_1-p,M_{\eta},M_{K^0})
\nonumber \\
&+& \left.4(2M_{K^{\pm}}^2-M_{\pi^{\pm}}^2-p\cdot p_1)B_{11}(p_1-p,M_{\pi^{\pm}},M_{K^{\pm}})\right\rbrace \,.
\nonumber
\end{eqnarray}

\subsection{The $u$-channel contribution}

Finally, the $u$-channel contribution follows from diagram (e) in figure~\ref{fig:strong},
\begin{eqnarray}
\delta F
&=& \delta G
\nonumber \\
&=& \dfrac{1}{6F_0^2}\left\lbrace A_0(M_{\pi^{\pm}})+A_0(M_{K^{\pm}})\right.
\nonumber \\
&+& \left.6(M_{\pi^{\pm}}^2-M_{\pi^0}^2+p\cdot p_2)B_0(p_2-p,M_{\pi^{\pm}},M_{K^{\pm}})\right\rbrace \,,
\nonumber
\end{eqnarray}

\subsection{Soft virtual photon contribution}

The various topologies of Feynman diagrams containing a virtual photon are drawn in figures~ \ref{fig:vertex-leg}, \ref{fig:leg-leg}, \ref{fig:R-vertex-leg}, and \ref{fig:R-leg-leg}. The contribution of each figure will be given separately. 

\subsubsection*{Figure~\ref{fig:vertex-leg} contribution}

\textbf{diagram 4. (a)}
\begin{equation}
\delta F\,=\,-\frac{8e^2}{3}\left\{ 2B_0(-p,0,M_K )+B_1(-p,0,M_K)\right\}\,, \quad
\delta G\,=\,0\,.
\end{equation}
\textbf{diagram 4. (b)}
\begin{equation}
\delta F\,=\,\delta G\,=\,0\,.
\end{equation}
\textbf{diagram 4. (c)}
\begin{equation}
\delta F\,=\,\delta G\,=\,-\frac{4e^2}{3}\left\{ 2B_0(-p_1,0,M_{\pi})+B_1(-p_1,0,M_{\pi})\right\}\,.
\end{equation}
\textbf{diagram 4. (d)}
\begin{equation}
\delta F\,=\,-\delta G\,=\,\frac{4e^2}{3}\left\{ 2B_0(-p_2,0,M_{\pi})+B_1(-p_2,0,M_{\pi})\right\}\,.
\end{equation}

\subsubsection*{Figure~\ref{fig:leg-leg} contribution}

\textbf{diagram 5. (a)}
\begin{eqnarray}
\delta F
&=& \delta G
\nonumber \\
&=& -e^2\left [ 4p\cdot p_2C_0(-p_2,-p,m_{\gamma},M_{\pi} ,M_K)\right.
\nonumber \\
&+& \left.B_0(-p_2,0,M_{\pi})+B_0(-p,0,M_K)-B_0(p_2-p,M_{\pi},M_K)\right ]\,.
\end{eqnarray}
\textbf{diagram 5. (b)}
\begin{eqnarray}
\delta F
&=& e^2\left [4p\cdot p_1C_0(-p_1,-p,m_{\gamma},M_{\pi} ,M_K )\right.
\nonumber \\
&+& B_0(-p_1,0,M_{\pi})+B_0(-p,0,M_K)-B_0(p_1-p,M_{\pi},M_K)
\nonumber \\
&+& 4p\cdot p_1C_1(-p_1,-p,0,M_{\pi},M_K)+8p\cdot p_1C_2(-p_1,-p,0,M_{\pi},M_K) \nonumber \\
&+& B_1(-p_1,0,M_{\pi})+2B_1(-p,0,M_K)
\nonumber \\
&+& \left.B_0(p_1-p,M_{\pi},M_K)-B_1(p_1-p,M_{\pi},M_K)\right ]
\nonumber \\
\delta G
&=& e^2\left [4p\cdot p_1C_0(-p_1,-p,m_{\gamma},M_{\pi},M_K)\right.
\nonumber \\
&+& B_0(-p_1,0,M_{\pi})+B_0(-p,0,M_K)-B_0(p_1-p,M_{\pi},M_K)
\nonumber \\
&+& B_0(p_1-p,M_{\pi},M_K)+B_1(p_1-p,M_{\pi},M_K)
\nonumber \\
&+& \left.B_1(-p_1,0,M_{\pi})+4p\cdot p_1C_1(-p_1,-p,0,M_{\pi},M_K)\right ]\,.
\end{eqnarray}
\textbf{diagram 5. (c)}
\begin{eqnarray}
\delta F
&=& \frac{e^2}{3}\left [6B_0(-p_l,0,m_l)\right.
\nonumber \\
&+& B_0(-p,0,M_K)+2B_1(-p,0,M_K)
\nonumber \\
&+& 12p\cdot p_lC_0(-p_l,-p,m_{\gamma},m_l,M_K)+2m_l^2C_1(-p_l,-p,0,m_l,M_K) \nonumber \\
&-& \left.6(M_K^2-2p\cdot p_l)C_2(-p_l,-p,0,m_l,M_K)\right ]\,,
\nonumber \\
\delta G
&=& \frac{e^2}{3}\left [3B_0(-p,0,M_K)\right.
\nonumber \\
&+& 12p\cdot p_lC_0(-p_l,-p,m_{\gamma},m_l,M_K)
\nonumber \\
&-& 6(m_l^2-2p\cdot p_l)C_1(-p_l,-p,0,m_l,M_K)
\nonumber \\
&+& \left.6M_K^2C_2(-p_l,-p,0,m_l,M_K)\right ]\,.
\end{eqnarray}
\textbf{diagram 5. (d)}
\begin{eqnarray}
\delta F
&=& -e^2\left [-B_0(p_1,0,M_{\pi})-B_0(-p_2,0,M_{\pi})\right.
\nonumber \\
&-& B_1(p_1,0,M_{\pi})+B_1(-p_2,0,M_{\pi})-2B_1(-p_1-p_2,M_{\pi},M_{\pi})
\nonumber \\
&+& 4p_1\cdot p_2C_0(p_1,-p_2,m_{\gamma},M_{\pi},M_{\pi})
\nonumber \\
&+& 4p_1\cdot p_2C_1(p_1,-p_2,0,M_{\pi},M_{\pi})
\nonumber \\
&-& \left.4p_1\cdot p_2C_2(p_1,-p_2,0,M_{\pi},M_{\pi})\right ]\,,
\nonumber \\
\delta G
&=& -e^2\left [-B_0(p_1,0,M_{\pi})-B_0(-p_2,0,M_{\pi})\right.
\nonumber \\
&-& B_1(p_1,0,M_{\pi})-B_1(-p_2,0,M_{\pi})
\nonumber \\
&+& 4p_1\cdot p_2C_0(p_1,-p_2,m_{\gamma},M_{\pi},M_{\pi})
\nonumber \\
&+& 4p_1\cdot p_2C_1(p_1,-p_2,0,M_{\pi},M_{\pi})
\nonumber \\
&+& \left.4p_1\cdot p_2C_2(p_1,-p_2,0,M_{\pi},M_{\pi})\right ]\,.
\end{eqnarray}
\textbf{diagram 5. (e)}
\begin{eqnarray}
\delta F
&=& \delta G
\nonumber \\
&=& \frac{e^2}{3}\left [-6B_0(p_l,0,m_l)\right.
\nonumber \\
&-& B_0(-p_1,0,M_{\pi})-2B_1(-p_1,0,M_{\pi})
\nonumber \\
&+& 12p_1\cdot p_lC_0(p_l,-p_1,m_{\gamma},m_l,M_{\pi})
\nonumber \\
&-& 2m_l^2C_1(p_l,-p_1,0,m_l,M_{\pi})
\nonumber \\
&+& \left.6(M_{\pi}^2+2p_1\cdot p_l)C_2(p_l,-p_1,0,m_l,M_{\pi})\right ]\,.
\end{eqnarray}
\textbf{diagram 5. (f)}
\begin{eqnarray}
\delta F
&=& -\frac{e^2}{3}\left [-6B_0(p_l,0,m_l)\right.
\nonumber \\
&-& B_0(-p_2,0,M_{\pi})+6B_0(-p_2-p_l,m_l,M_{\pi})+B_1(-p_2,0,M_{\pi})
\nonumber \\
&+& 12p_2\cdot p_lC_0(p_l,-p_2,m_{\gamma},m_l,M_{\pi})
\nonumber \\
&-& 2m_l^2C_1(p_l,-p_2,0,m_l,M_{\pi})
\nonumber \\
&+& \left.6M_{\pi}^2C_2(p_l,-p_2,0,m_l,M_{\pi})\right ]\,,
\nonumber \\
\delta G
&=& -\frac{e^2}{3}\left [-6B_0(p_l,0,m_l)\right.
\nonumber \\
&-& 5B_0(-p_2,0,M_{\pi})+6B_0(-p_2-p_l,m_l,M_{\pi})-B_1(-p_2,0,M_{\pi})
\nonumber \\
&+& 12p_2\cdot p_lC_0(p_l,-p_2,m_{\gamma},m_l,M_{\pi})
\nonumber \\
&+& 2m_l^2C_1(p_l,-p_2,0,m_l,M_{\pi})
\nonumber \\
&+& \left.6M_{\pi}^2C_2(p_l,-p_2,0,m_l,M_{\pi})\right ]\,.
\end{eqnarray}

\subsubsection*{Figure~\ref{fig:R-vertex-leg} contribution}

\textbf{diagram 6. (a)}
\begin{eqnarray}
\delta F
&=& -\frac{e^2}{6}\left [-2B_0(-p_l-p_{\nu},0,M_K)-2B_0(-p,0,M_K)-B_1(-p,0,M_K)\right.
\nonumber \\
&+& 12(M_{\pi}^2+p_1\cdot p_2-p\cdot p_1+p\cdot p_2)C_0(-p,-p_l-p_{\nu},0,M_K,M_K) \nonumber \\
&+& 6(M_{\pi}^2+p_1\cdot p_2-3p\cdot p_1+3p\cdot p_2)C_1(-p,-p_l-p_{\nu},0,M_K,M_K)
\nonumber \\
&-& 12(p\cdot p_1-p\cdot p_2)C_2(-p,-p_l-p_{\nu},0,M_K,M_K)
\nonumber \\
&-& 6(p\cdot p_1-p\cdot p_2)C_{11}(-p,-p_l-p_{\nu},0,M_K,M_K)
\nonumber \\
&-& \left.6(p\cdot p_1-p\cdot p_2)C_{12}(-p,-p_l-p_{\nu},0,M_K,M_K)\right ]\,, \nonumber \\
\delta G
&=& e^2C_{00}(-p,-p_l-p_{\nu},0,M_K,M_K)\,.
\end{eqnarray}
\textbf{diagram 6. (b)}
\begin{equation}
\delta F\,=\,\delta G\,=\,-\frac{4e^2}{3}\, B_0(-p_l-p_{\nu},0,M_K)\,.
\end{equation}
\textbf{diagram 6. (c)}
\begin{eqnarray}
\delta F
&=& \delta G
\nonumber \\
&=& \frac{4e^2}{3}\left [B_0(-p_l-p_{\nu},0,M_K)\right.
\nonumber \\
&-& m_l^2C_1(-p_l-p_{\nu},-p_l,0,M_K,m_l)
\nonumber \\
&-& \left.m_l^2C_2(-p_l-p_{\nu},-p_l,0,M_K,m_l)\right ]\,.
\end{eqnarray}
\textbf{diagram 6. (d)}
\begin{eqnarray}
\delta F
&=& \delta G
\nonumber \\
&=& -\frac{e^2}{12}\left [4B_0(-p_l-p_{\nu},0,M_K)+4B_0(p_1,0,M_{\pi})+2B_1(p_1,0,M_{\pi})\right.
\nonumber \\
&+& 6(M_K^2+2M_{\pi}^2-m_l^2
\nonumber \\
&& +2p_1\cdot p_2-2p\cdot p_1+2p\cdot p_2-2p_l\cdot p_{\nu})C_0(p_1,-p_l-p_{\nu},0,M_{\pi},M_K)
\nonumber \\
&+& 3(M_K^2+2M_{\pi}^2-m_l^2+2p_1\cdot p_2
\nonumber \\
&& \left.-2p\cdot p_1+2p\cdot p_2-2p_l\cdot p_{\nu})C_1(p_1,-p_l-p_{\nu},0,M_{\pi},M_K) \right ]\,.
\end{eqnarray}
\textbf{diagram 6. (e)}
\begin{eqnarray}
\delta F
&=& \frac{e^2}{6}\left [2B_0(p_2,0,M_{\pi})-4B_0(-p_l-p_{\nu},0,M_K)+B_1(p_2,0,M_{\pi})\right.
\nonumber \\
&+& 3(M_K^2+2M_{\pi}^2-m_l^2
\nonumber \\
&& +2p_1\cdot p_2-2p\cdot p_1+2p\cdot p_2-2p_l\cdot p_{\nu})C_0(p_2,-p_l-p_{\nu},0,M_{\pi},M_K)
\nonumber \\
&+& 18p\cdot p_2C_1(p_2,-p_l-p_{\nu},0,M_{\pi},M_K)
\nonumber \\
&-& 6(M_K^2-2M_{\pi}^2+m_l^2-2p_1\cdot p_2+2p_l\cdot p_{\nu})C_2(p_2,-p_l-p_{\nu},0,M_{\pi},M_K)
\nonumber \\
&+& 12C_{00}(p_2,-p_l-p_{\nu},0,M_{\pi},M_K)
\nonumber \\
&+& 6p\cdot p_2C_{11}(p_2,-p_l-p_{\nu},0,M_{\pi},M_K)
\nonumber \\
&-& 3(M_K^2-2M_{\pi}^2+m_l^2
\nonumber \\
&& \left.-2p_1\cdot p_2+2p_l\cdot p_{\nu})C_{12}(p_2,-p_l-p_{\nu},0,M_{\pi},M_K)\right ]\,,
\nonumber \\
\delta G
&=& -\frac{e^2}{6}\left [2B_0(p_2,0,M_{\pi})-4B_0(-p_l-p_{\nu},0,M_K)+B_1(p_2,0,M_{\pi})\right.
\nonumber \\
&+& 3(M_K^2+2M_{\pi}^2-m_l^2
\nonumber \\
&& +2p_1\cdot p_2-2p\cdot p_1+2p\cdot p_2-2p_l\cdot p_{\nu})C_0(p_2,-p_l-p_{\nu},0,M_{\pi},M_K)
\nonumber \\
&+&  18p\cdot p_2C_1(p_2,-p_l-p_{\nu},0,M_{\pi},M_K)
\nonumber \\
&-& 6(M_K^2-2M_{\pi}^2+m_l^2-2p_1\cdot p_2+2p_l\cdot p_{\nu})C_2(p_2,-p_l-p_{\nu},0,M_{\pi},M_K)
\nonumber \\
&+& 6p\cdot p_2C_{11}(p_2,-p_l-p_{\nu},0,M_{\pi},M_K)
\nonumber \\
&-& 3(M_K^2-2M_{\pi}^2+m_l^2
\nonumber \\
&& \left.-2p_1\cdot p_2+2p_l\cdot p_{\nu})C_{12}(p_2,-p_l-p_{\nu},0,M_{\pi},M_K)\right ]\,.
\nonumber
\end{eqnarray}

\subsubsection*{Figure~\ref{fig:R-leg-leg} contribution}

\textbf{diagram 7. (a)}
\begin{eqnarray}
\delta F
&=& \frac{e^2}{6}\left\{-2B_0(-p_l-p_{\nu},0,M_K)-2B_0(-p,0,M_K)-B_1(-p,0,M_K)\right. \nonumber \\
&+& 12(M_{\pi}^2+p_1\cdot p_2-p\cdot p_1+p\cdot p_2)C_0(-p,-p_l-p_{\nu},0,M_K,M_K) \nonumber \\
&-& 12(p\cdot p_1-p\cdot p_2)C_1(-p,-p_l-p_{\nu},0,M_K,M_K)
\nonumber \\
&+& 6(M_{\pi}^2+p_1\cdot p_2-p\cdot p_1+p\cdot p_2)C_1(-p,-p_l-p_{\nu},0,M_K,M_K)
\nonumber \\
&+& 2m_l^2C_1(-p_l,-p_l-p_{\nu},0,m_l,M_K)
\nonumber \\
&-& 12(p\cdot p_1-p\cdot p_2)C_2(-p,-p_l-p_{\nu},0,M_K,M_K)
\nonumber \\
&+& 2m_l^2C_2(-p_l,-p_l-p_{\nu},0,m_l,M_K)
\nonumber \\
&+& 2m_l^2C_2(-p,-p_l,0,M_K,m_l)
\nonumber \\
&-& 6(p\cdot p_1-p\cdot p_2)C_{11}(-p,-p_l-p_{\nu},0,M_K,M_K)
\nonumber \\
&-& 6(p\cdot p_1-p\cdot p_2)C_{12}(-p,-p_l-p_{\nu},0,M_K,M_K)
\nonumber \\
&-& 12m_l^2(M_{\pi}^2+p_1\cdot p_2-p\cdot p_1+p\cdot p_2)\times
\nonumber \\
&& \qquad D_2(-p,-p_l,-p_l-p_{\nu},0,M_K,m_l,M_K)
\nonumber \\
&-& 12m_l^2(M_{\pi}^2+p_1\cdot p_2-p\cdot p_1+p\cdot p_2)\times
\nonumber \\
&& \qquad D_3(-p,-p_l,-p_l-p_{\nu},0,M_K,m_l,M_K)
\nonumber \\
&+& 12m_l^2(p_1\cdot p_l-p_2\cdot p_l)\times
\nonumber \\
&& \qquad D_{22}(-p,-p_l,-p_l-p_{\nu},0,M_K,m_l,M_K)
\nonumber \\
&+& 12m_l^2(p\cdot p_1-p\cdot p_2)\times
\nonumber \\
&& \qquad D_{33}(-p,-p_l,-p_l-p_{\nu},0,M_K,m_l,M_K)  \nonumber \\
&+& 12m_l^2(p\cdot p_1-p\cdot p_2)\times
\nonumber \\
&& \qquad D_{12}(-p,-p_l,-p_l-p_{\nu},0,M_K,m_l,M_K)  \nonumber \\
&+& 12m_l^2(p\cdot p_1-p\cdot p_2)\times
\nonumber \\
&& \qquad D_{13}(-p,-p_l,-p_l-p_{\nu},0,M_K,m_l,M_K)  \nonumber \\
&+& 12m_l^2(p\cdot p_1-p\cdot p_2+p_1\cdot p_l-p_2\cdot p_l)\times
\nonumber \\
&& \qquad \left.D_{23}(-p,-p_l,-p_l-p_{\nu},0,M_K,m_l,M_K)\right\}\,,
\nonumber \\
\delta G
&=& -e^2\left [C_{00}(-p,-p_l-p_{\nu},0,M_K,M_K)\right.
\nonumber \\
&+& \left.2m_l^2D_{00}(-p,-p_l,-p_l-p_{\nu},0,M_K,m_l,M_K)\right ]\,.
\end{eqnarray}
\textbf{diagram 7. (b)}
\begin{eqnarray}
\delta F
&=& \delta G
\nonumber \\
&=& \frac{e^2}{6}\left\{2B_0(-p_l-p_{\nu},0,M_K)\right.
\nonumber \\
&+& 2B_0(p_1,0,M_{\pi})+B_1(p_1,0,M_{\pi})
\nonumber \\
&+& 12p\cdot p_2C_0(p_1,-p_l-p_{\nu},0,M_{\pi},M_K)
\nonumber \\
&+& 6p\cdot p_2C_1(p_1,-p_l-p_{\nu},0,M_{\pi},M_K)
\nonumber \\
&-& 2m_l^2C_1(-p_l,-p_l-p_{\nu},0,m_l,M_K)
\nonumber \\
&-& 2m_l^2C_2(-p_l,-p_l-p_{\nu},0,m_l,M_K)
\nonumber \\
&-& 2m_l^2C_2(p_1,-p_l,0,M_{\pi},m_l)
\nonumber \\
&-& 12m_l^2p\cdot p_2D_2(p_1,-p_l,-p_l-p_{\nu},0,M_{\pi},m_l,M_K)
\nonumber \\
&-& \left.12m_l^2p\cdot p_2D_3(p_1,-p_l,-p_l-p_{\nu},0,M_{\pi},m_l,M_K) \right\}\,.
\end{eqnarray}
\textbf{diagram 7. (c)}
\begin{eqnarray}
\delta F
&=& -\frac{e^2}{6}\left\{-4B_0(p_2,0,M_{\pi})\right.
\nonumber \\
&+& 2B_0(-p_l-p_{\nu},0,M_K)-2B_1(p_2,0,M_{\pi})
\nonumber \\
&-& 6(M_K^2-m_l^2
\nonumber \\
&& -2p\cdot p_2-2p_l\cdot p_{\nu})C_0(p_2,-p_l-p_{\nu},0,M_{\pi},M_K) \nonumber \\
&-& 3(M_K^2-m_l^2
\nonumber \\
&& -4p_1\cdot p_2-2p\cdot p_2-2p_l\cdot p_{\nu})C_1(p_2,-p_l-p_{\nu},0,M_{\pi},M_K)
\nonumber \\
&-& 2m_l^2C_1(-p_l,-p_l-p_{\nu},0,m_l,M_K)
\nonumber \\
&+& 12(M_{\pi}^2+p_1\cdot p_2-p\cdot p_1)C_2(p_2,-p_l-p_{\nu},0,M_{\pi},M_K)
\nonumber \\
&+& 4m_l^2C_2(p_2,-p_l,0,M_{\pi},m_l)
\nonumber \\
&-& 2m_l^2C_2(-p_l,-p_l-p_{\nu},0,m_l,M_K) \nonumber \\
&+& 6C_{00}(p_2,-p_l-p_{\nu},0,M_{\pi},M_K)
\nonumber \\
&+& 6p_1\cdot p_2C_{11}(p_2,-p_l-p_{\nu},0,M_{\pi},M_K)
\nonumber \\
&+& 6(M_{\pi}^2+p_1\cdot p_2-p\cdot p_1)C_{12}(p_2,-p_l-p_{\nu},0,M_{\pi},M_K)
\nonumber \\
&+& 6m_l^2(M_K^2-m_l^2-2p\cdot p_2-2p_l\cdot p_{\nu})\times
\nonumber \\
&& \qquad D_2(p_2,-p_l,-p_l-p_{\nu},0,M_{\pi},m_l,M_K)
\nonumber \\
&+& 6m_l^2(M_K^2-m_l^2-2p\cdot p_2-2p_l\cdot p_{\nu})\times
\nonumber \\
&& D_3(p_2,-p_l,-p_l-p_{\nu},0,M_{\pi},m_l,M_K)
\nonumber \\
&+& 12m_l^2p_1\cdot p_lD_{22}(p_2,-p_l,-p_l-p_{\nu},0,M_{\pi},m_l,M_K)
\nonumber \\
&-& 12m_l^2(M_{\pi}^2+p_1\cdot p_2-p\cdot p_1)\times
\nonumber \\
&& \qquad D_{33}(p_2,-p_l,-p_l-p_{\nu},0,M_{\pi},m_l,M_K)
\nonumber \\
&-& 12m_l^2p_1\cdot p_2D_{12}(p_2,-p_l,-p_l-p_{\nu},0,M_{\pi},m_l,M_K)
\nonumber \\
&-& 12m_l^2p_1\cdot p_2D_{13}(p_2,-p_l,-p_l-p_{\nu},0,M_{\pi},m_l,M_K)
\nonumber \\
&-& 12m_l^2(M_{\pi}^2+p_1\cdot p_2-p\cdot p_1-p_1\cdot p_l)\times
\nonumber \\
&& \qquad \left.D_{23}(p_2,-p_l,-p_l-p_{\nu},0,M_{\pi},m_l,M_K)\right\}\,,
\nonumber \\
\delta G
&=& \frac{e^2}{6}\left\{-4B_0(p_2,0,M_{\pi})\right.
\nonumber \\
&+& 2B_0(-p_l-p_{\nu},0,M_K)-2B_1(p_2,0,M_{\pi})
\nonumber \\
&-& 6(M_K^2-m_l^2
\nonumber \\
&& -2p\cdot p_2-2p_l\cdot p_{\nu})C_0(p_2,-p_l-p_{\nu},0,M_{\pi},M_K) \nonumber \\
&-& 3(M_K^2-m_l^2
\nonumber \\
&& -4p_1\cdot p_2-2p\cdot p_2-2p_l\cdot p_{\nu})C_1(p_2,-p_l-p_{\nu},0,M_{\pi},M_K)
\nonumber \\
&-& 2m_l^2C_1(-p_l,-p_l-p_{\nu},0,m_l,M_K)
\nonumber \\
&+& 12(M_{\pi}^2+p_1\cdot p_2-p\cdot p_1)C_2(p_2,-p_l-p_{\nu},0,M_{\pi},M_K)
\nonumber \\
&+& 4m_l^2C_2(p_2,-p_l,0,M_{\pi},m_l)
\nonumber \\
&-& 2m_l^2C_2(-p_l,-p_l-p_{\nu},0,m_l,M_K)
\nonumber \\
&-& 6C_{00}(p_2,-p_l-p_{\nu},0,M_{\pi},M_K)
\nonumber \\
&+& 6p_1\cdot p_2C_{11}(p_2,-p_l-p_{\nu},0,M_{\pi},M_K)
\nonumber \\
&+& 6(M_{\pi}^2+p_1\cdot p_2-p\cdot p_1)C_{12}(p_2,-p_l-p_{\nu},0,M_{\pi},M_K)
\nonumber \\
&+& 6m_l^2(M_K^2-m_l^2-2p\cdot p_2-2p_l\cdot p_{\nu})\times
\nonumber \\
&& \qquad D_2(p_2,-p_l,-p_l-p_{\nu},0,M_{\pi},m_l,M_K)
\nonumber \\
&+& 6m_l^2(M_K^2-m_l^2-2p\cdot p_2-2p_l\cdot p_{\nu})\times
\nonumber \\
&& \qquad D_3(p_2,-p_l,-p_l-p_{\nu},0,M_{\pi},m_l,M_K)
\nonumber \\
&+& 12m_l^2p_1\cdot p_lD_{22}(p_2,-p_l,-p_l-p_{\nu},0,M_{\pi},m_l,M_K)
\nonumber \\
&-& 12m_l^2(M_{\pi}^2+p_1\cdot p_2-p\cdot p_1)\times
\nonumber \\
&& \qquad D_{33}(p_2,-p_l,-p_l-p_{\nu},0,M_{\pi},m_l,M_K)
\nonumber \\
&-& 12m_l^2p_1\cdot p_2D_{12}(p_2,-p_l,-p_l-p_{\nu},0,M_{\pi},m_l,M_K)
\nonumber \\
&-& 12m_l^2p_1\cdot p_2D_{13}(p_2,-p_l,-p_l-p_{\nu},0,M_{\pi},m_l,M_K)
\nonumber \\
&-& 12m_l^2(M_{\pi}^2+p_1\cdot p_2-p\cdot p_1-p_1\cdot p_l)\times
\nonumber \\
&& \qquad \left.D_{23}(p_2,-p_l,-p_l-p_{\nu},0,M_{\pi},m_l,M_K)\right\}\,.
\end{eqnarray}

%% file: section_3.tex
\section{Soft photon bremsstrahlung}

Virtual photon corrections to $K_{\ell 4}$ decay rate generate
infrared divergencies. These cancel, order by order in
perturbation theory, with the ones coming from real bremsstrahlung
corrections. Assume that the emitted photons are soft, that is,
their energies are smaller than any detector resolution, $\omega$.
It follows that radiative and non-radiative decays cannot be
distinguished experimentally and emission of real soft photons
should be taken into account. Note however that only single soft
photon radiation is needed to one-loop accuracy.

\subsection{The decay amplitude}

A general feature of photon bremsstrahlung is that, in the soft
photon approximation, the bremsstrahlung amplitude is proportional
to the Born amplitude. Since we deal only with isospin breaking
corrections to the $F$ and $G$ form factors, the Born amplitude is
taken, all along this section, to be,
$$
{\cal A}_{\mathrm{B}}^{+-}\,=\,\frac{1}{2F_0}\,G_FV_{us}^*{\bar
u}(p_{\nu})\gamma_{\mu}(1-\gamma^5)v(p_l)(P^{\mu}+Q^{\mu})\,.
$$
The contribution of form factors $F$ and $G$ to the Bremsstrahlung
amplitude can be read off from diagrams in figure~\ref{fig:soft}.

Let $\varepsilon$ and $q$ be, respectively, the polarization
vector and the momentum of the radiated photon. The evaluation of
diagrams in figure~\ref{fig:soft} is straightforward and read, to first
order in the photon energy,
\begin{equation} \label{eq:bremsstrahlung
amplitude} {\cal A}^{+-\gamma}\,=\,e{\cal
A}_{\mathrm{B}}^{+-}\left (-\frac{p\cdot\varepsilon^*}{p\cdot
q}+\frac{p_l\cdot\varepsilon^*}{p_l\cdot
q}+\frac{p_1\cdot\varepsilon^*}{p_1\cdot
q}-\frac{p_2\cdot\varepsilon^*}{p_2\cdot q}\right )\,. \end{equation}

Squaring the matrix element (\ref{eq:bremsstrahlung amplitude})
and summing over polarizations, we obtain
\begin{eqnarray}
\sum_{\mathrm{pol.}}|{\cal A}^{+-\gamma}|^2
  &=& -e^2|{\cal A}_{\mathrm{B}}^{+-}|^2\times \nonumber \\
  & & \left [\frac{M_K^2}{(p\cdot q)^2}+\frac{m_l^2}{(p_l\cdot q)^2}
        +\frac{M_{\pi}^2}{(p_1\cdot q)^2}+\frac{M_{\pi}^2}{(p_2\cdot
        q)^2}\right.
        \nonumber \\
  &-& \frac{2p\cdot p_1}{(p\cdot q)(p_1\cdot q)}+
        \frac{2p\cdot p_2}{(p\cdot q)(p_2\cdot q)}-
        \frac{2p\cdot p_l}{(p\cdot q)(p_l\cdot q)} \nonumber \\
  &+& \left.\frac{2p_1\cdot p_l}{(p_1\cdot q)(p_l\cdot q)}-
        \frac{2p_2\cdot p_l}{(p_2\cdot q)(p_l\cdot q)}-
        \frac{2p_1\cdot p_2}{(p_1\cdot q)(p_2\cdot q)}\right ]\,.
        \nonumber
\end{eqnarray}

The preceding expression is singular for vanishing momentum of the
soft photon. We shall attribute a small but non-vanishing mass to
the photon, $m_{\gamma}$, in order to regularize this singularity.

\subsection{The decay rate}

The $K_{\ell 4\gamma}$ differential decay rate is obtained by
squaring the matrix element (\ref{eq:bremsstrahlung amplitude}),
summing over spins and polarizations and integrating over the
photon phase space,
\begin{eqnarray}
 d\Phi_{\gamma}
  &=& (2\pi )^4\delta^{(4)}(p_1+p_2+p_l+p_{\nu}+q-p)\times \nonumber \\
  & & \frac{d^3\boldsymbol{p}_1}{(2\pi )^32E_1}\,
        \frac{d^3\boldsymbol{p}_2}{(2\pi )^32E_2}\,
        \frac{d^3\boldsymbol{p}_l}{(2\pi )^32E_l}\,
        \frac{d^3\boldsymbol{p}_{\nu}}{(2\pi
        )^32|\boldsymbol{p}_{\nu}|}\,\frac{d^3\boldsymbol{q}}{(2\pi )^32|\boldsymbol{q}|}\,.
        \nonumber
\end{eqnarray}
Using the definition of bremsstrahlung integrals as given in the
appendix of reference~\cite{Nehme:2003bz}, the $K_{\ell 4\gamma}$ differential decay rate takes the following form in the soft photon approximation,
\begin{eqnarray}
d\Gamma_{\gamma}
&=& \frac{1}{2M_K}\,d\Phi\left (-\frac{e^2}{2}\right )\sum_{\mathrm{spins}}|{\cal A}_{\mathrm{B}}^{+-}|^2\bigg [\,I(p,p,m_{\gamma},\omega )
\nonumber \\
&+& I(p_l,p_l,m_{\gamma},\omega )
+I(p_1,p_1,m_{\gamma},\omega )+I(p_2,p_2,m_{\gamma},\omega )
\nonumber \\
&-& 2I(p,p_1,m_{\gamma},\omega )
+2I(p,p_2,m_{\gamma},\omega )-2I(p,p_l,m_{\gamma},\omega )
\nonumber \\
&+& 2I(p_1,-p_l,m_{\gamma},\omega )
-2I(p_2,-p_l,m_{\gamma},\omega )-2I(p_1,-p_2,m_{\gamma},\omega )\,\bigg ]\,,
\label{eq:decay_rate}
\end{eqnarray}
where non-singular $m_{\gamma}$ terms have been dropped out.

\subsection{Cancellation of infrared divergencies}

The infrared divergent part of $K_{\ell 4\gamma}$ differential
decay rate can be extracted from (\ref{eq:decay_rate}) using the
definition of bremsstrahlung integrals from the appendix of reference~\cite{Nehme:2003bz},
\begin{eqnarray}
d\Gamma_{\gamma}^{\mathrm{IR}}
  &=& \frac{e^2}{4\pi^2}\,\frac{1}{2M_K}\,d\Phi\,\sum_{\mathrm{spins}}|{\cal A}_{\mathrm{B}}^{+-}|^2\times \nonumber \\
  & & \bigg [\,2+p_1\cdot p_2\tau (p_1,-p_2,M_{\pi} ,M_{\pi} ) \nonumber \\
  & & -p\cdot p_1\tau (p,p_1,M_K ,M_{\pi} )+p\cdot p_2\tau (p,p_2,M_K ,M_{\pi} )
\nonumber \\ 
&& -p\cdot p_l\tau (p,p_l,M_K ,m_l)-p_1\cdot p_l\tau (p_1,-p_l,M_{\pi} ,m_l)
\nonumber \\ 
&& +p_2\cdot p_l\tau (p_2,-p_l,M_{\pi},m_l)\,\bigg ]\,\ln m_{\gamma}^2\,. \nonumber
\end{eqnarray}
On the other hand, the infrared divergence coming from virtual
photon corrections to $F$ and $G$ form factors only can be read
off from table~\ref{tab:infrared} and is denoted by $d\Gamma^{\mathrm{IR}}$. It is easy then to check that infrared divergencies cancel at the level of differential decay rates,
\begin{equation}
d\Gamma^{\mathrm{IR}}+d\Gamma_{\gamma}^{\mathrm{IR}}\,=\,0\,.
\end{equation}

%% file: conclusion.tex
\section{Perspectives}
\label{sec:perspectives} 

In this work we studied the decay process, $K^+\rightarrow\pi^+\pi^-\ell^+\nu_\ell$, taking into account Isospin breaking effects. These come mainly from electroweak interactions and generate corrections proportional to the fine structure constant, $\alpha$, and to the difference between up and down quark masses, $m_u-m_d$. 

The interest in this decay comes from the fact that the partial wave expansion of the corresponding form factors involves $\pi\pi$ scattering phase shifts. The latter can be related in a model-independent way to the $\pi\pi$ scattering lengths which are sensitive to the value of the quark condensate. Thus, a precise measurement of form factors should allow accurate determination of scattering lengths and, consequently, give precious information about the QCD vacuum structure. 

Scattering lengths are strong interaction quantities. On the other hand, any $K_{\ell 4}$ decay measurement contains contributions from all possible interactions, in particular, from electroweak ones. Therefore, it is primordial to have under control Isospin breaking effects in order to disentangle the strong interaction contribution from the measured form factors. The present work was guided by this motivation and, to this end, analytic expressions for form factors were obtained including Isospin breaking effects. These expressions are ultraviolet finite, scale independent, but infrared divergent. We showed that this divergence cancels out at the differential decay rate level if we take into account real soft photon emission. 

Our work should be completed by, 
\begin{itemize}
\item a parametrization of form factors in the presence of Isospin breaking,
\item a full treatment of the radiative decay, $K_{\ell 4\gamma}$.  
\end{itemize} 
Calculations in this direction are in progress. 

%% file: paper.bbl
\begin{thebibliography}{10}

\bibitem{Stern:1998dy}
Jan Stern.
\newblock Two alternatives of spontaneous chiral symmetry breaking in {QCD}.
\newblock 1998.

\bibitem{Gell-Mann:1968rz}
Murray Gell-Mann, R.~J. Oakes, and B.~Renner.
\newblock Behavior of current divergences under su(3) x su(3).
\newblock {\em Phys. Rev.}, 175:2195--2199, 1968.

\bibitem{Knecht:1995tr}
M.~Knecht, B.~Moussallam, J.~Stern, and N.~H. Fuchs.
\newblock The low-energy pi pi amplitude to one and two loops.
\newblock {\em Nucl. Phys.}, B457:513--576, 1995.

\bibitem{Watson:1952ji}
Kenneth~M. Watson.
\newblock The effect of final state interactions on reaction cross- sections.
\newblock {\em Phys. Rev.}, 88:1163--1171, 1952.

\bibitem{Watson:1954uc}
Kenneth~M. Watson.
\newblock Some general relations between the photoproduction and scattering of
  pi mesons.
\newblock {\em Phys. Rev.}, 95:228--236, 1954.

\bibitem{Ananthanarayan:2000ht}
B.~Ananthanarayan, G.~Colangelo, J.~Gasser, and H.~Leutwyler.
\newblock Roy equation analysis of pi pi scattering.
\newblock {\em Phys. Rept.}, 353:207--279, 2001.

\bibitem{Efimov:1986}
G.~V. Efimov, M.~A. Ivanov, and V.~E. Lyubovitski\u{\i}.
\newblock On the characteristics of the $\pi^+\pi^-$ atom.
\newblock {\em Sov. J. Nucl. Phys.}, 44:296--299, 1986.

\bibitem{Adeva:2003up}
B.~Adeva et~al.
\newblock Dirac: A high resolution spectrometer for pionium detection.
\newblock 2003.

\bibitem{Pislak:2001bf}
S.~Pislak et~al.
\newblock A new measurement of k+(e4) decay and the s-wave pi pi scattering
  length a(0)(0).
\newblock {\em Phys. Rev. Lett.}, 87:221801, 2001.

\bibitem{Pislak:2003sv}
S.~Pislak et~al.
\newblock High statistics measurement of k(e4) decay properties.
\newblock {\em Phys. Rev.}, D67:072004, 2003.

\bibitem{Colangelo:2001sp}
G.~Colangelo, J.~Gasser, and H.~Leutwyler.
\newblock The quark condensate from k(e4) decays.
\newblock {\em Phys. Rev. Lett.}, 86:5008--5010, 2001.

\bibitem{Descotes-Genon:2001tn}
S.~Descotes-Genon, N.~H. Fuchs, L.~Girlanda, and J.~Stern.
\newblock Analysis and interpretation of new low-energy pi pi scattering data.
\newblock {\em Eur. Phys. J.}, C24:469--483, 2002.

\bibitem{Batley:2000}
R.~Batley et~al. NA48~Collaboration.
\newblock For a precision measurement of charged kaon decay parameters with an
  extended na48 setup.
\newblock Approved proposal CERN/SPSC 2000-003, CERN, 2000.

\bibitem{Santos:2003}
E.~Santos. KTeV~Collaboration.
\newblock An analysis of
  $k_l\longrightarrow\pi^0\pi^{\mp}\mathrm{e}^{\pm}\nu_{\mathrm{e}}(\overline{%
\nu}_{\mathrm{e}})$.
\newblock Talk given at the apr03 meeting of the american physical society,
  Philadelphia, April 5-8, 2003.

\bibitem{Maltman:1997nw}
Kim Maltman and Carl~E. Wolfe.
\newblock Electromagnetic corrections to pi pi scattering lengths: Some lessons
  for the implementation of meson exchange models.
\newblock {\em Phys. Lett.}, B393:19--25, 1997.

\bibitem{Maltman:1997nwE}
Kim Maltman and Carl~E. Wolfe.
\newblock Electromagnetic corrections to pi pi scattering lengths: Some lessons
  for the implementation of meson exchange models.
\newblock {\em Phys. Lett.}, B424:413, 1998.

\bibitem{Meissner:1997fa}
Ulf-G. Meissner, G.~Muller, and S.~Steininger.
\newblock Virtual photons in su(2) chiral perturbation theory and
  electromagnetic corrections to pi pi scattering.
\newblock {\em Phys. Lett.}, B406:154--160, 1997.

\bibitem{Meissner:1997faE}
Ulf-G. Meissner, G.~Muller, and S.~Steininger.
\newblock Virtual photons in su(2) chiral perturbation theory and
  electromagnetic corrections to pi pi scattering.
\newblock {\em Phys. Lett.}, B407:454, 1997.

\bibitem{Knecht:1998jw}
Marc Knecht and Res Urech.
\newblock Virtual photons in low energy pi pi scattering.
\newblock {\em Nucl. Phys.}, B519:329--360, 1998.

\bibitem{Knecht:2002gz}
M.~Knecht and A.~Nehme.
\newblock Electromagnetic corrections to charged pion scattering at low
  energies.
\newblock {\em Phys. Lett.}, B532:55--62, 2002.

\bibitem{Nehme:2003bz}
A.~Nehme.
\newblock Isospin breaking in k(l4) decays of the neutral kaon.
\newblock 2003.

\bibitem{Callan:1966hu}
C.~G. Callan and S.~B. Treiman.
\newblock Equal time commutators and k meson decays.
\newblock {\em Phys. Rev. Lett.}, 16:153--157, 1966.

\bibitem{Weinberg:1966}
S.~Weinberg.
\newblock Current-commutator calculation of the $k_{\ell 4}$ form factors.
\newblock {\em Phys. Rev. Lett.}, 17:336--340, 1966.

\bibitem{Weinberg:1966E}
S.~Weinberg.
\newblock Current-commutator calculation of the $k_{\ell 4}$ form factors.
\newblock {\em Phys. Rev. Lett.}, 18:1178, 1967.

\bibitem{Berman:1968ss}
S.~M. Berman and Probir Roy.
\newblock Soft pion theorems and the k(l3), k(l4) form-factors.
\newblock {\em Phys. Lett.}, B27:88--91, 1968.

\bibitem{Cabibbo:1965}
N.~Cabibbo and A.~Maksymowicz.
\newblock Angular correlations in $k_{\mathrm{e}4}$ decays and determination of
  low-energy $\pi$-$\pi$ phase shifts.
\newblock {\em Phys. Rev.}, 137:B438--B443, 1965.

\bibitem{Knecht:1999ag}
M.~Knecht, H.~Neufeld, H.~Rupertsberger, and P.~Talavera.
\newblock Chiral perturbation theory with virtual photons and leptons.
\newblock {\em Eur. Phys. J.}, C12:469--478, 2000.

\bibitem{Coleman:1969sm}
Sidney~R. Coleman, J.~Wess, and Bruno Zumino.
\newblock Structure of phenomenological lagrangians. 1.
\newblock {\em Phys. Rev.}, 177:2239--2247, 1969.

\bibitem{Callan:1969sn}
Jr. Callan, Curtis~G., Sidney~R. Coleman, J.~Wess, and Bruno Zumino.
\newblock Structure of phenomenological lagrangians. 2.
\newblock {\em Phys. Rev.}, 177:2247--2250, 1969.

\bibitem{Gasser:1985gg}
J.~Gasser and H.~Leutwyler.
\newblock Chiral perturbation theory: Expansions in the mass of the strange
  quark.
\newblock {\em Nucl. Phys.}, B250:465, 1985.

\bibitem{Urech:1995hd}
Res Urech.
\newblock Virtual photons in chiral perturbation theory.
\newblock {\em Nucl. Phys.}, B433:234--254, 1995.

\bibitem{Neufeld:1995eg}
H.~Neufeld and H.~Rupertsberger.
\newblock Isospin breaking in chiral perturbation theory and the decays eta
  $\to$ pi lepton neutrino and tau $\to$ eta pi neutrino.
\newblock {\em Z. Phys.}, C68:91--102, 1995.

\bibitem{Neufeld:1996mu}
H.~Neufeld and H.~Rupertsberger.
\newblock The electromagnetic interaction in chiral perturbation theory.
\newblock {\em Z. Phys.}, C71:131--138, 1996.

\end{thebibliography}
